\documentclass[superscriptaddress,twocolumn,prx,aps,preprintnumbers,notitlepage,longbibliography,floatfix]{revtex4-1}

\usepackage[latin9]{inputenc}
\usepackage{amsmath}
\usepackage{amssymb}
\usepackage{graphicx}
\usepackage{esint}
\usepackage[scr=boondoxo, scrscaled=1.05]{mathalfa}

\usepackage{censor}

\usepackage{hyperref}
\allowdisplaybreaks
\usepackage{ulem}
\makeatletter

\pdfpageheight\paperheight
\pdfpagewidth\paperwidth

\pdfoutput=1

\usepackage{epsfig}
\usepackage{graphics}
\graphicspath{{figures/}}

\usepackage{mciteplus}
\mciteErrorOnUnknownfalse

\makeatother
\usepackage{color}

\begin{document}
\StopCensoring

\title{Angular correlations on  causally-coherent inflationary horizons}


\author{Craig Hogan} 
\affiliation{University of Chicago, 5640 South Ellis Ave., Chicago, IL 60637}
\author{Stephan S. Meyer}
\affiliation{University of Chicago, 5640 South Ellis Ave., Chicago, IL 60637}
\author{Nathaniel Selub}
\affiliation{University of Chicago, 5640 South Ellis Ave., Chicago, IL 60637}

\author{Frederick Wehlen}
\affiliation{University of Chicago, 5640 South Ellis Ave., Chicago, IL 60637}
\begin{abstract}
We develop a  model for correlations of  cosmic microwave background (CMB) anisotropy on the largest angular scales, based on standard causal geometrical relationships in  slow-roll inflation.  Unlike standard models based on quantized field modes, it  describes perturbations with  nonlocal directional  coherence on spherical boundaries of causal diamonds. Causal constraints reduce the number of independent degrees of freedom, impose new  angular symmetries, and eliminate cosmic variance for purely angular 2-point correlations. Distortions of  causal structure from vacuum fluctuations are modeled as gravitational memory from randomly oriented outgoing and incoming  gravitational null shocks, with nonlocally coherent directional displacements on curved surfaces of causal diamonds formed by  standard inflationary horizons. The angular distribution is determined by  axially symmetric shock displacements  on circular intersections of the comoving sphere that represents the CMB photosphere with other inflationary horizons--- those centered on it, and those that pass through an observer's world line.  Displacements on thin spheres at the end of inflation have a unique angular power spectrum $C_\ell$ that approximates the standard expectation on small angular scales, but differs substantially at large angular scales due to horizon curvature.  For a thin sphere, the model  predicts a  universal angular correlation function $C(\Theta)$ with an exact ``causal shadow'' symmetry,  $C(\pi/4<\Theta<3\pi/4)= 0$, and significant large-angle parity violation.   We apply a rank statistic to compare models with  {\sl WMAP} and {\sl Planck} satellite data, and find that a causally-coherent model with no shape parameters or cosmic variance  agrees with the measured $C(\Theta)$ better than a large fraction ($> 0.9999$) of standard model realizations. Model-independent tests of holographic causal symmetries are proposed.

\end{abstract}

\maketitle

\section{Introduction}

In  relativistic cosmological models, the pattern of  structure on the largest scales survives intact from  the earliest epochs, with minimal influence from subsequent cosmic evolution \cite{1967ApJ...147...73S}. 
Thus, measurements of  large-angle anisotropy in the cosmic microwave background (CMB), which provide direct comparisons of physical quantities on the largest scales, provide
the most  precise probe we have of relationships among the earliest events.  
For example, the measured angular correlation function $C(\Theta)$ of CMB temperature perturbation $\delta T$, or all-sky average product of $\delta T$ at points with angular separation $\Theta$, was already largely fixed by initial gravitational perturbations  
 on a large thin sphere in cosmological initial conditions at the end of the early inflationary era.

Ever since the first measurements of primordial anisotropy in the  CMB \cite{1992ApJ...396L..13W,1994ApJ...436..423B,Hinshaw_1996},
it has been remarked  that  $C(\Theta)$ at large angles is surprisingly small.
Indeed, 
in all-sky maps generated from the {\sl WMAP} and {\sl Planck} satellite data \cite{WMAPanomalies,Ade:2015hxq,Akrami:2019bkn}, after allowing for the fact that the intrinsic dipole is degenerate with the motion of the solar system relative to the local cosmic rest frame,   $C(\Theta)$ appears to be not just remarkably small, but actually consistent with zero,  over a significant range of angular separation near 90 degrees \cite{Hagimoto_2020,Hogan_2022}.

In the standard quantum model for the origin of perturbations,
the
actual $C(\Theta)$ is  one  of many possibilities, almost all of which depart from zero at large angular scales much more than the real sky.  In
that
framework, the  measured small value of $C(\Theta)$ is simply a statistical fluke \cite{WMAPanomalies,Ade:2015hxq,Akrami:2019bkn,deOliveira-Costa:2003utu,Schwarz:2015cma}, and  
large-angle anisotropy is often disregarded.

In this paper, we analyze and test the hypothesis that small correlations in the large-angle CMB pattern are not an accident, but
are a consequence of  new and profound precise  angular symmetries of cosmic initial conditions  that  follow if
 primordial  gravitational perturbations  arise from quantum states that created directionally coherent displacements on the causal diamonds defined by a standard inflationary background\cite{Hogan2019,Hogan2020pattern,Hogan_2022}.
 This  ``causal coherence'' is not a property of quantum states in the standard  model for the origin of  inflationary perturbations based on vacuum fluctuations of  field modes, but we argue below 
that it  would be expected for gravitational fluctuations in relational and holographic theories of quantum gravity and locality.

 The present study extends our previous work with a  causally-coherent geometrical model of angular structure in inflationary perturbations.
We construct a  model of  holographic distortions of causal structure in a standard classical slow-roll inflation background, assembled from relational displacements  produced by coherent addition of  virtual
gravitational null shocks on horizons of every world line. The model leads to a definite    $C(\Theta)$ for causal distortion on thin spheres at the end of inflation with no cosmic variance.
We compare the  model with the current CMB measurements, and find that it matches the real sky much better than  the standard model at large angular separations.
At the same time, it reproduces the  standard mean inflationary power spectrum on smaller scales, with the intriguing addition of significant universal parity violation.
We suggest that measured  symmetries of CMB anisotropy at large angular separation could provide  direct evidence for  holographic causal   coherence  of quantum gravity.

\begin{figure*}[hbt]
 \centering
 \includegraphics[width=.9\textwidth]{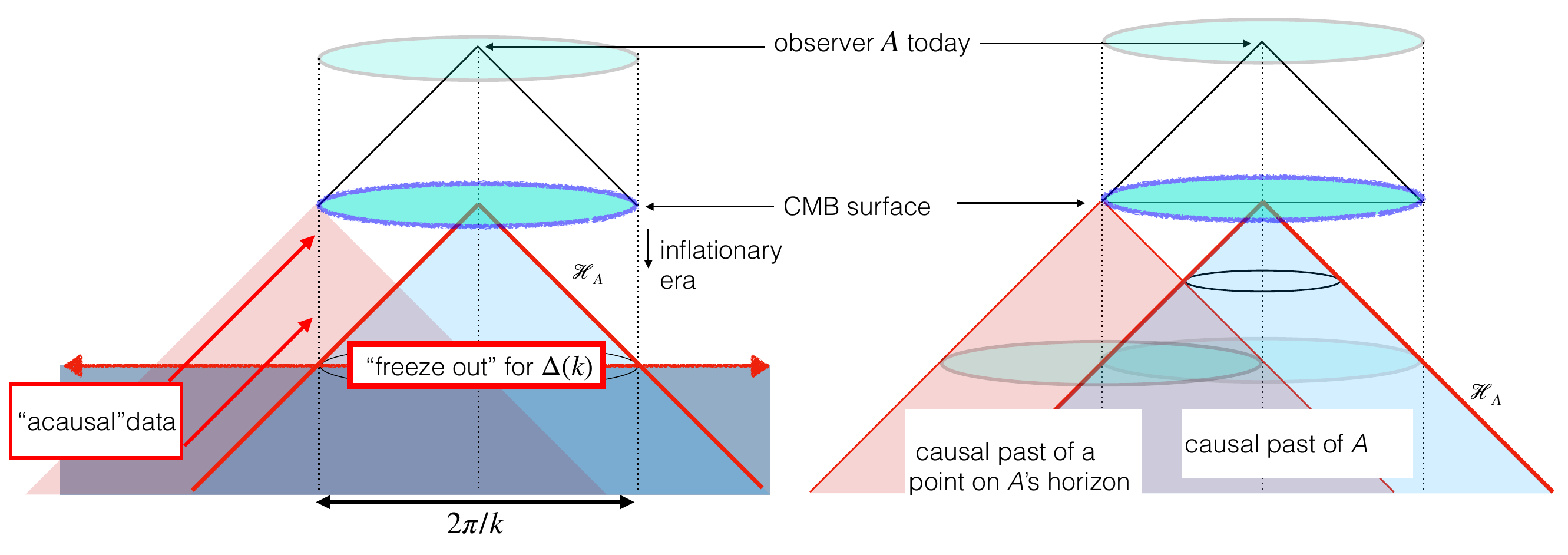}
 \caption{Space-time diagrams of an inflationary universe, in conformal comoving coordinates, on a logarithmic scale. The current horizon of world line $A$ intersects the CMB sky on a comoving sphere, or horizon footprint, that exited the inflationary horizon ${\cal H}_A$ during inflation roughly 60 $e$-foldings before it ended.
 The left panel illustrates standard quantum inflation based on QFT. Quantum coherence is assigned to plane wave modes of an acausally prepared initial vacuum state.  A classical value of metric distortion $\Delta_k$ on scale $k$ freezes out coherently on surfaces of constant  time as  comoving spacelike plane waves, which requires acausal spatial  coherence of  modes out to spatial infinity in the planar directions. At right, in our holographic    model of  fluctuations,  anisotropy is generated by causally prepared,   directionally coherent  displacements on  inflationary horizons  where  quantum fluctuations freeze into classical perturbations.
 \label{causality}}
\end{figure*}

 
 \section{Causally-coherent geometrical fluctuations}

\subsection{Locality and causal coherence}

In general the preparation, reduction and collapse of a quantum state, and its emergence as an approximately classical system, is a completely delocalized process. Thus, the spatial effect of a quantum-state reduction depends on model assumptions about the spatial coherence of system states.
To be consistent with causality, the preparation and  completion of a quantum measurement by any observer must lie within a causal diamond of that observer.  We will refer to this property of a system as ``causal coherence.'' 


The  formation of primordial perturbations from quantum fluctuations of the cosmic vacuum during inflation\cite{Baumann:2009ds} is based on 
 a  standard approach to connect classical space-time with a quantum system, 
the linearized quantum field theory of gravity (QFT).
It is widely thought that this approach is valid as long as inflation occurs well below the Planck energy scale. However, as explained below, it is not causally coherent.

The QFT model system has spatially infinite coherent wave states. Its vacuum states, which are obtained by creation operators on spatially infinite modes, are completely spatially delocalized. These are the states whose zero-point fluctuations give rise to cosmic fluctuations. In a relational quantum theory of  locality, fluctuations of a completely delocalized state should have no observable effect.

For a sum of wavelike perturbations to be causally coherent with a sharply defined classical  causal diamond,  waves with different frequencies and directions cannot have independent phases.
The  transform of a  sharp edge such as a causal horizon is a phase-coherent superposition of all frequencies; a sharp spherical horizon around a point entangles all frequencies and directions.
As discussed below in the context of the Einstein-Podolsky-Rosen (EPR) system, causally coherent  gravity  entangles  energy and geometry of all modes within 4D causal diamonds. The same coherence should apply to  fluctuations of vacuum states.

Indeed, in the presence of active gravity,  inconsistencies of QFT on large length scales are well known\cite{HollandsWald2004,Stamp_2015}.
This infrared problem can be traced to the renormalization needed to formulate QFT, which
does not correctly account for gravitational entanglement of quantum states with large-scale causal structure. As described by Hollands and Wald\cite{HollandsWald2004},
because of ``the holistic nature of renormalization theory, . . . an individual mode will have no way of knowing whether its own subtraction is correct unless it `knows' how the subtractions are being done for all other modes.''

 \subsection{Causal coherence in inflation}

 The only domain where  we  actually measure the frozen pattern from the active gravity of a virtual fluctuation  is in the pattern of primordial cosmic perturbations from quantum fluctuations during inflation.
 In standard quantum inflation theory based on QFT,  the pattern is determined by the initial conditions assumed for the quantum vacuum, which  adopts independent  acausal primordial phases in the initial state (Fig. \ref{causality}). 
 If the real system is causally coherent, almost  all  of these realizations are unphysical.

Although some cosmological  consequences of QFT infrared problems have been thoroughly studied\cite{CohenKaplanNelson1999,Banks2020}, they have not been incorporated into standard inflation theory.
In the standard description of cosmological ``measurement,'' or reduction of a quantum state to a classical system, vacuum fluctuations freeze into classical perturbations during inflation when their oscillation rate falls below the expansion rate. The QFT modes then ``collapse'' into definite classical amplitudes and phases that depend on initial conditions. Each mode is spatially infinite, so its state is not causally prepared; its spatial coherence is simply posited as an initial condition. 
In particular, as discussed below,   acausally prepared  long-wavelength  modes typically produce significant CMB correlations at large angular separations that are not observed.

In principle, quantum coherence is causal by construction in holographic or thermodynamic theories of gravity, which are built out of coherent causal diamonds\cite{Jacobson1995,Jacobson:2015hqa}. 
Coherent states of horizons and causal diamonds have been studied formally for anti-de Sitter space\cite{Ryu:2006bv,Verlinde:2019ade},
black holes \cite{Hooft:2016cpw,Hooft:2016itl,Hooft2018,Giddings:2018koz,Giddings:2019vvj,Haco:2018ske,Almheiri:2020cfm},
 early-universe cosmology\cite{Banks:2011qf,Banks:2015iya,Banks:2018}, and
flat space-time\cite{Verlinde:2019xfb,BanksZurek2021,BanksZurek2021b}. Holographic theories appear to be a consistent approach to quantum gravity, but as yet there is no theoretical consensus about specific new observable consequences, such as correlations of holographic quantum fluctuations. They have been sought in laboratory experiments that have achieved better than Planck scale precision\cite{holoshear,Richardson:2020snt}, but have not yet been detected. 
 
 Here, we estimate observable consequences of causal coherence for inflationary perturbations from a classical model of angular noise designed to emulate the causal coherence of a holographic theory of quantum gravity consistent with causally coherent emergent locality.
 Our semiclassical model does not yet connect directly with formal holographic quantum theories, but it is based on the same covariant causal principles. It allows us to generate specific holographic patterns that can be tested with
 real-world data: 
 the model provides realizations of cosmic perturbations on a spherical horizon footprint, which approximately correspond to CMB temperature maps on large angular scales. Statistical properties of the model can be compared with the real sky, and with  realizations of standard inflationary quantum theory on large angular scales.

\subsection{Causally coherent fluctuations }

In QFT, the states of the gravitational vacuum fluctuate like other fields. Each spatially-infinite mode has an amplitude quantized like a harmonic oscillator with zero-point fluctuations. In standard inflation theory, the modes are quantized on the accelerating expanding background, and their zero-point oscillations get ``frozen in'' to the classical metric on infinite spacelike surfaces. 
As noted above, this model is not causally coherent. 
This feature makes a significant difference for the angular structure of quantum fluctuations whose states are  reduced on spherical causal surfaces, the inflationary horizons of world lines.

Here, we adopt an alternative approach to model   the angular distribution of vacuum fluctuations of  causally coherent quantum gravity. The main difference  from the standard model is that  fluctuations  are modeled as  coherent directional displacements on horizons, instead of spatially infinite coherent plane waves.

 \begin{figure*}[hbt]
 \centering
 \includegraphics[width=.8\textwidth]{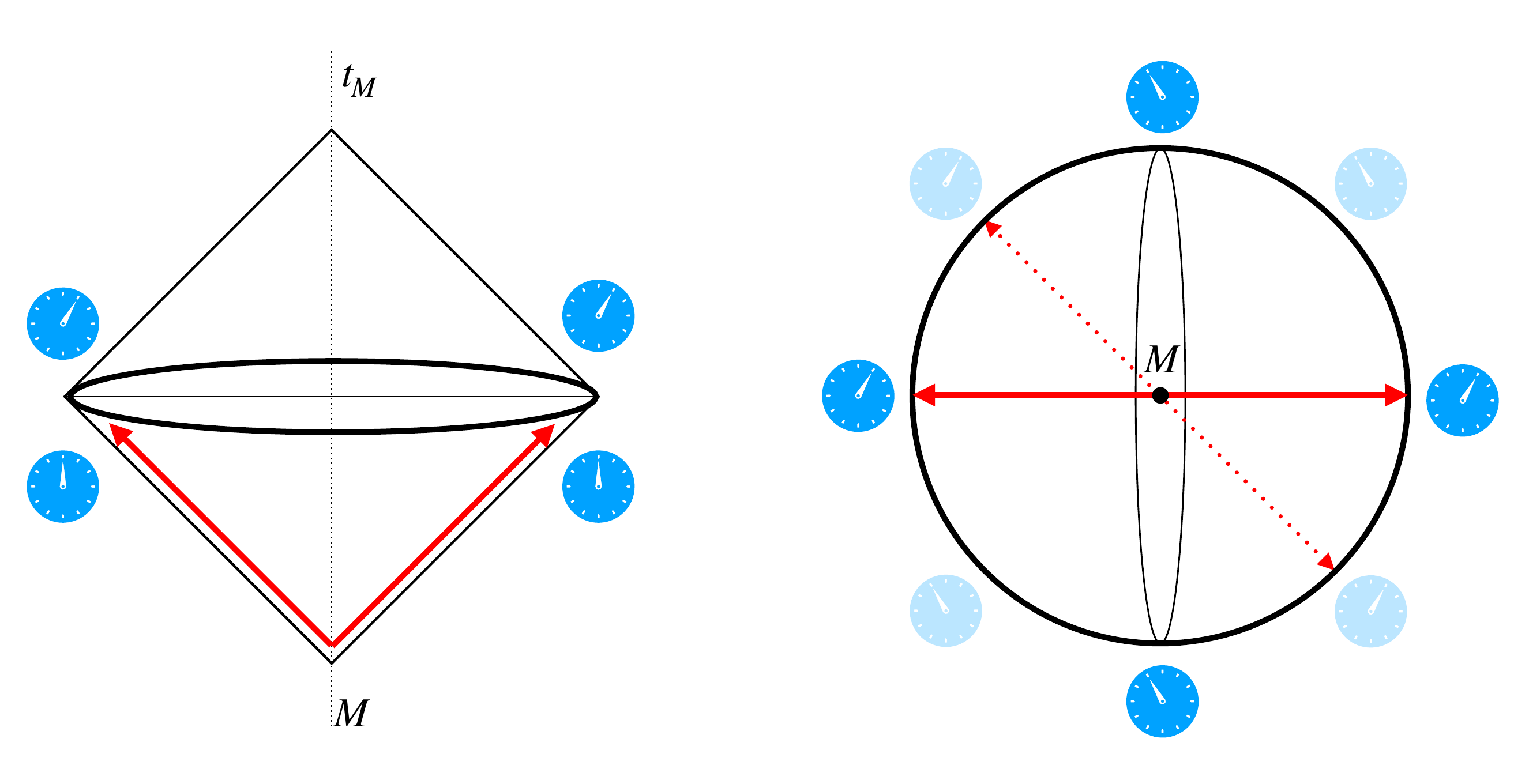}
 \caption{Angular structure of the gravitational shock wave generated by the Einstein-Podolsky-Rosen (EPR) system,  a particle of mass  $M$ that decays into a photon pair\cite{Mackewicz2022}. The left panel shows the causal diamond formed by the outgoing  pair, and incoming light used by an observer at the center to measure displacements on a sphere of clocks. The right panel  shows a spatial slice  of the clocks on the surface of the causal diamond, with
 the axially symmetric anisotropic displacement  generated by
the spherical null shock aligned with the decay axis. 
The  angular spectrum of clock displacements $\delta \tau(\theta,\phi)$ can be computed analytically in linear theory\cite{Mackewicz2022}, and is independent of sphere radius (Eq. \ref{timedisplace}).
Most of the total displacement occurs at large angles (Eq. \ref{ell2pattern}). 
 In a quantum EPR process, the decay  axis, and therefore the geometry, is in a coherent quantum superposition that encompasses the whole causal diamond; different axes create different patterns of displacement, as shown by shaded clocks for an alternative axis. Our model incorporates similar macroscopic coherence into causal diamonds of an inflationary background geometry.
 \label{clocks}}
\end{figure*}

 \subsubsection{Causally coherent gravitational shocks in EPR}
 
A good starting point for modeling causally coherent displacements is a classical gravitational null shock\cite{Aichelburg1971,DRAY1985,Tolish:2014a,Satishchandran:2019,Mackewicz2022}.
 A null point particle, such as a photon of momentum $ p$, creates a  discontinuous position displacement 
\begin{equation}\label{displacement}
  \delta x =  (G/c^3)  \   p .
\end{equation}
The observable physical effect depends on the  position and motion of the observer in relation to the photon trajectory.
 In general, it leaves a directional ``memory'' of displacements in relation to a focal point: radial positions, or  times measured by clocks, are displaced in an axially symmetric pattern 
 \begin{equation}
      c \delta \tau = \delta x = d(\theta)  
 \end{equation}
 where $\theta$ denotes the angle from the photon axis. 
 Our model posits similar 
 macroscopic directional coherence for coherent virtual gravitational fluctuations.

A recently analyzed classical example of a curved shock with anisotropic displacements is the spherical gravitational shock wave from the EPR system, the two-photon decay of a massive particle\cite{Mackewicz2022}. 
The projection of the displacement memory from the gravitational shock creates a large-angle pattern of time distortions  on clocks viewed on the surface of a causal diamond in the frame of the decaying particle (Fig. \ref{clocks}). 

The angular pattern of remembered time displacements can be solved analytically in linear theory\cite{Mackewicz2022}, with  a spectral decomposition into axially symmetric $(m=0)$ spherical harmonics:
\begin{equation}\label{timedisplace}
    \delta\tau(\theta,\phi)= \frac{2GM}{c^3}\sum_{\ell=2,{\rm even}}^{\infty} \frac{2\ell+1}{\ell(\ell-1)(\ell+1)(\ell+2)} P_\ell ({\rm cos}\theta),
\end{equation}
where $M$ denotes the total mass and  $P_\ell$ are Legendre polynomials.

This function describes the permanent residual memory of large-angle distortions of causal relationships in relation to a particular point, the center of a causal diamond. Notably, the displacement does not depend on the size of the causal diamond: at any separation, it is  macroscopically coherent, with a total displacement  dominated by the lowest harmonics. For example, the quadrupolar component of   displacement $d_{\ell=2}(\theta)$ at an angle $\theta$ from the decay axis, viewed from the center, is comparable to the total displacement  (Eq. \ref{displacement}):
\begin{equation}\label{ell2pattern}
 \delta\tau_2\equiv d_{\ell=2}(\theta)/c= \frac{5}{24}\frac{GM }{c^3}(3 {\rm cos}^2\theta-1).
\end{equation}
 Although it is a real physical distortion of causal structure, the radial displacement does not carry energy.
Large-angle directional coherence is a generic feature of remembered causal distortion patterns  caused by  passage of  gravitational shocks.

The right side of Fig. \ref{clocks} illustrates the same solution applied to a quantum EPR system with an indeterminate decay axis.  The geometry of a causal diamond of any size around the decay is placed into a causally coherent superposition with  large-angle distortion on the surface  given by Eq. (\ref{timedisplace}). 
In the 3D spectrum of this quantum system, pointlike particles with momenta along a single axis  have wave functions with entangled axially symmetric momentum components in all directions, and at all wavelengths, to
localize the state to the axis. The same entanglement must apply to the wave function of geometry.
Similar   causally-coherent anisotropic displacements of causal structure by virtual spherical null shocks on inflationary horizons are used here to make a model of gravitational displacement noise on spheres at the end of inflation.

\subsubsection{Causally coherent fluctuations from virtual null shocks}

The quantum degrees of freedom of a causally coherent theory are not the same as the classical macroscopic ones; it is said that
 classical space-time, and locality in space and time, emerge statistically or ``holographically'' from a quantum system. 
 Although holographic emergence of classical space-time has been extensively studied\cite{Stamp_2015}, including some models that incorporate virtual shocks and holography into quantum models of horizons\cite{Hooft:2018syc,BanksZurek2021,Liu_2022,Verlinde2022}, there is no standard theory of their phenomenology. 
Our semiclassical angular-noise model is designed to allow estimates of such holographic effects in real-world data. The goal is to test whether the pattern of fluctuations in a causally-coherent geometrical quantum vacuum could have distinctive large-angle symmetries that resemble those of the real sky.

We  use displacements from virtual gravitational shocks  to build a causally coherent model of inflationary fluctuations. 
This model of fluctuations  has physically significant macroscopic differences from QFT. First, their displacements are not scalars, but like the gravitational shocks in EPR they leave a gravitational memory with an associated direction.
Second, their causal coherence is different: a quantized plane wave depends on information at spacelike infinity in all directions, whereas spherical null shocks  create coherent causal directional displacements associated with causal diamonds of particular world lines. 
A null-shock quantum model requires macroscopic coherence (and wavefunction ``collapse'') everywhere on a null surface, which leads to ``spooky'', albeit causal, nonlocal spacelike quantum correlations, like the  EPR system. 
 As described above in that system, shock displacements are causally coherent at large angles, even on macroscopic scales.
The large-scale correlations of distortions in these two models are very different, a reflection of the IR entanglement of long-wavelength modes with causal structure produced by gravity and neglected in QFT\cite{Stamp_2015,HollandsWald2004}.

 \subsubsection{Magnitude of coherent perturbations on horizons}

Suppose that  virtual geometrical shocks create coherent displacements on the bounding surfaces of  causal diamonds with zero mean, and produce individual displacements in causal structure with a magnitude given by the classical value (Eqs. \ref{displacement}, \ref{ell2pattern}). 
In a  causal diamond of duration $\tau= R/c$, the surface   fluctuates  with displacement $d$ given by the sum of the virtual displacements of many virtual shocks, so it has a variance given by the sum of the individual variances.

Suppose further that in a causally-coherent,  holographic theory of gravity,   virtual shocks arrive with frequency and variance given by the Planck time $t_P$.  The fractional distortion  $\Delta$  then  has variance\cite{Hogan2012,Hogan:2016,Mackewicz2022}
\begin{equation}\label{displacementvariance}
\langle d^2\rangle / R^2 \simeq \langle\Delta^2\rangle\simeq ct_P/R.
\end{equation}
Because of the large-scale coherence of the distortions, this is much larger than the fluctuation produced by QFT fluctuations, which are of the order of  $(ct_P/R)^2$. In principle, such large coherent distortions of causal diamonds in flat space-time could be detected in interferometer experiments\cite{holoshear,Richardson:2020snt,VERLINDE2021136663,Kwon2022,Li2022,McCuller2022}.
 
When applied to the curved space-time of slow-roll inflation, the same idea leads to perturbation power per $e$-folding of comoving length scale ${\cal R}$ given by
the physical horizon radius in Planck units\cite{Hogan2019,Hogan2020pattern,Hogan_2022}:
\begin{equation}\label{potentialvariance}
 d \langle \Delta^2 \rangle/ d\ln {\cal R} \simeq H t_P,
\end{equation}
where $H= \dot a/ a$ is the expansion rate of the cosmic scale factor $a$. This quantum-gravitational variance is again typically many orders of magnitude larger than that predicted by the QFT model, which are of the order of $(H t_P)^2$, multiplied by the inverse of a slow-roll parameter $\epsilon$ that depends on the slope of the inflaton potential\cite{Hogan2019}. 

Thus, even though an inflationary potential can be chosen for either model that matches the data, in any given  background the gravity of vacuum inflaton fluctuations, which is the main effect in the standard picture, is generally negligible compared to coherent quantum-gravity fluctuations in causal structure.
This difference ultimately results from the foundational differences in the space-time structure of coherent states that follows from the model of locality adopted for the quantum system.


\section{Causally-coherent model of inflationary perturbations}

\subsection{Causal structure of standard inflation}

Our  model of  fluctuations is based on the classical causal structure of standard slow-roll inflation.
 An unperturbed inflationary 
universe has a 
Friedmann-Lema\^itre-Robertson-Walker metric, with space-time interval 
\begin{equation}\label{FLRW}
ds^2 = a^2(t) [c^2d\eta^2- d\Sigma^2],
\end{equation}
where $t$ denotes proper cosmic time for any comoving observer, $d\eta\equiv dt/ a(t)$ denotes a conformal time interval, and $a(t)$ denotes the cosmic scale factor, determined by the equations of motion and a model for the inflaton potential. 
The spatial 3-metric in comoving coordinates is
\begin{equation}\label{flatspace}
d\Sigma^2 = dr^2 + r^2 d\Omega^2,
\end{equation}
where the angular interval in standard polar notation is $d\Omega^2 = d\theta^2 + \sin^2 \theta d\phi^2$.
Future and past light cones from an event are defined by a null path,
\begin{equation}\label{null}
d\Sigma = \pm cd\eta,
\end{equation}
so conformal causal structure is the same as in flat space-time. 
A causal diamond with boundary at $t$ corresponds to an interval with equal conformal time before and after $t$.

\begin{figure}[t]
\begin{centering}
\includegraphics[width=\linewidth]{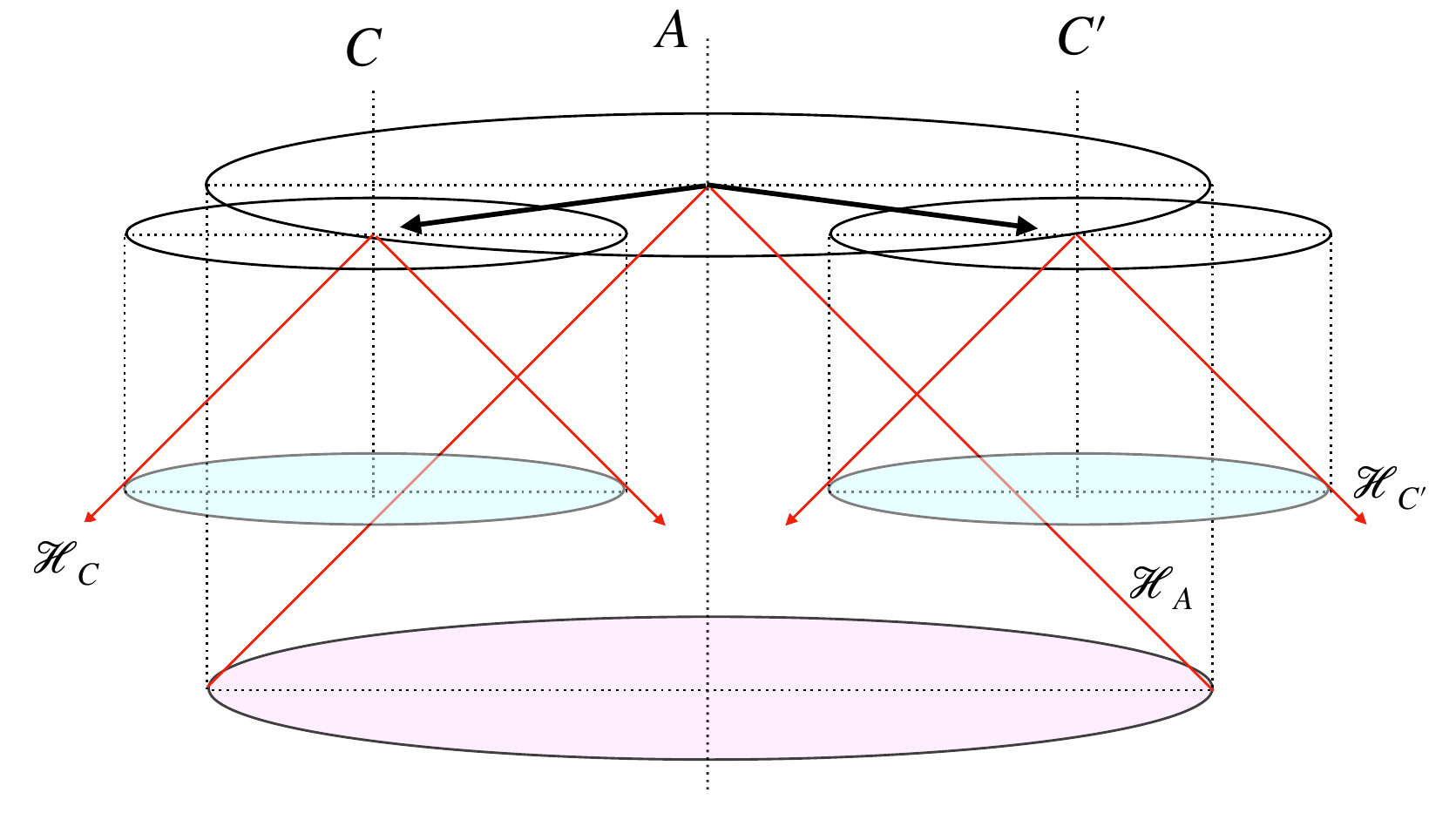}
\par\end{centering}
\protect\caption{ Causal history of anisotropy observed from world line $A$. Vertical lines represent constant comoving positions.  The top surface represents a surface of constant time at the end of  inflation, which on this scale coincides with the epoch of CMB last scattering.
The  past light cone of $A$ represents its inflationary horizon
${\cal H}_A$.
The large circle at the top represents $A$'s current horizon,  a thin sphere that approximates the
 CMB last scattering surface and represents
 the comoving surface of a causal diamond defined by the present epoch at $A$'s location. Quantum fluctuations  during inflation produce classical perturbations on comoving ``footprints'' where world-cylinders intersect horizons.
Two  points $C,C'$ on the $A$ horizon footprint are shown as examples, at a large angular separation in directions shown  by  arrows, with their respective  horizons ${\cal H}_C$ and
${\cal H}_{C'}$.
In the causally coherent model, anisotropy  generated by  quantum fluctuations  is modeled as displacements by null virtual shocks on horizons. 
\label{AChorizons}}
\end{figure}

The classical inflationary horizon of any comoving world line is an incoming null surface that arrives   at the end of inflation.
During the slow roll phase of  inflation, the 
horizon has an approximately constant physical radius $c/H$, determined by the expansion rate $H$.
 A comoving sphere  passes through the horizon at a particular time, and represents the``horizon footprint'' of that time
 (Fig. \ref{AChorizons}).

In a causally-coherent model\cite{Hogan_2022},   correlations in CMB anisotropy are generated by directional displacements from fluctuations of  coherent quantum objects,  causal diamonds whose boundaries are determined by inflationary horizons.
As in standard inflation, relational quantum fluctuations get frozen into classical perturbations when they cross horizon surfaces.  Our model tracks  nonlocal directional correlations of causal relationships,  taking into account the curvature of the  null surfaces.


\subsection{Angular relationships of intersecting horizons}

\begin{figure*}[t]
\begin{centering}
\includegraphics[width=.45\linewidth]{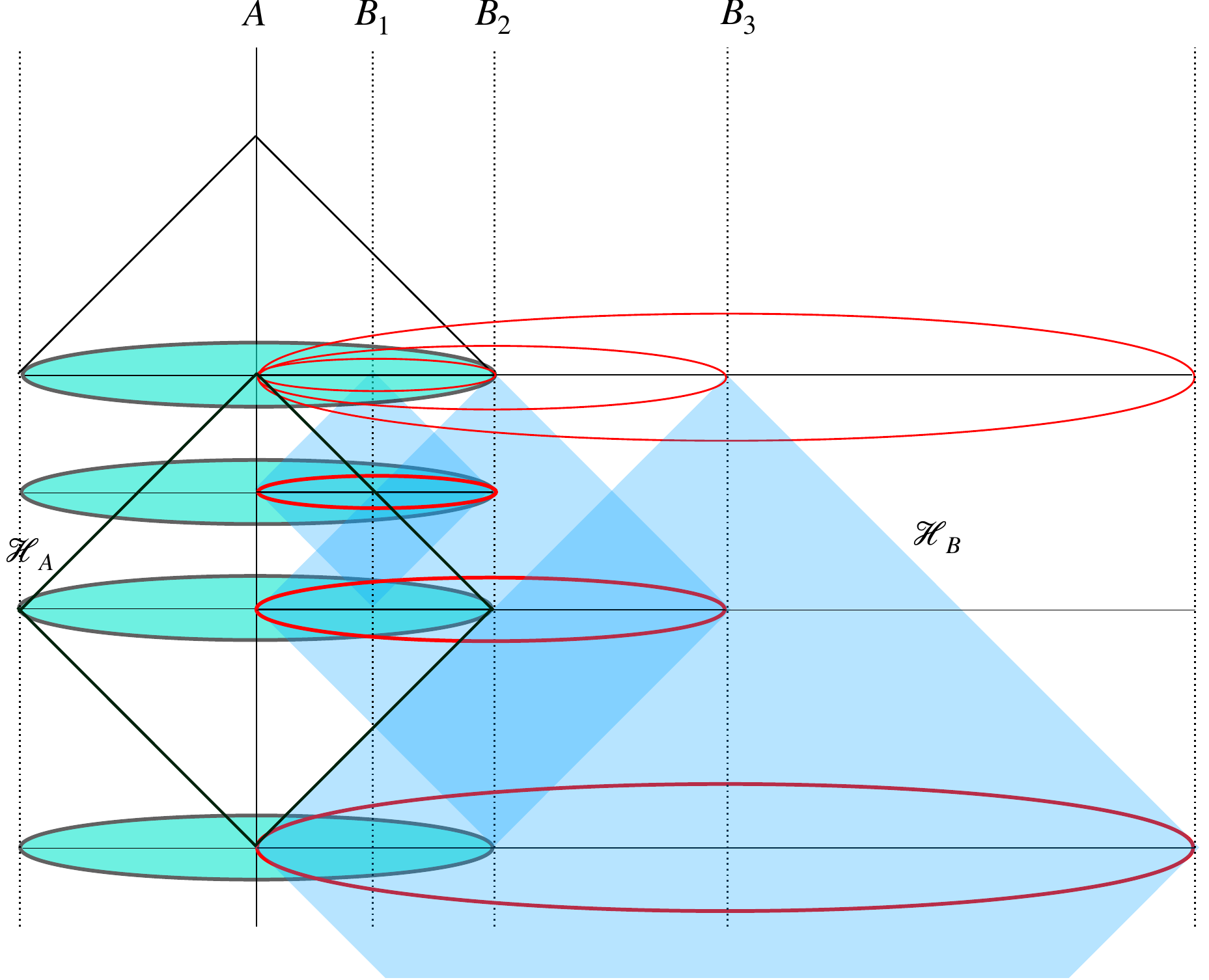}
\includegraphics[width=.49\linewidth]{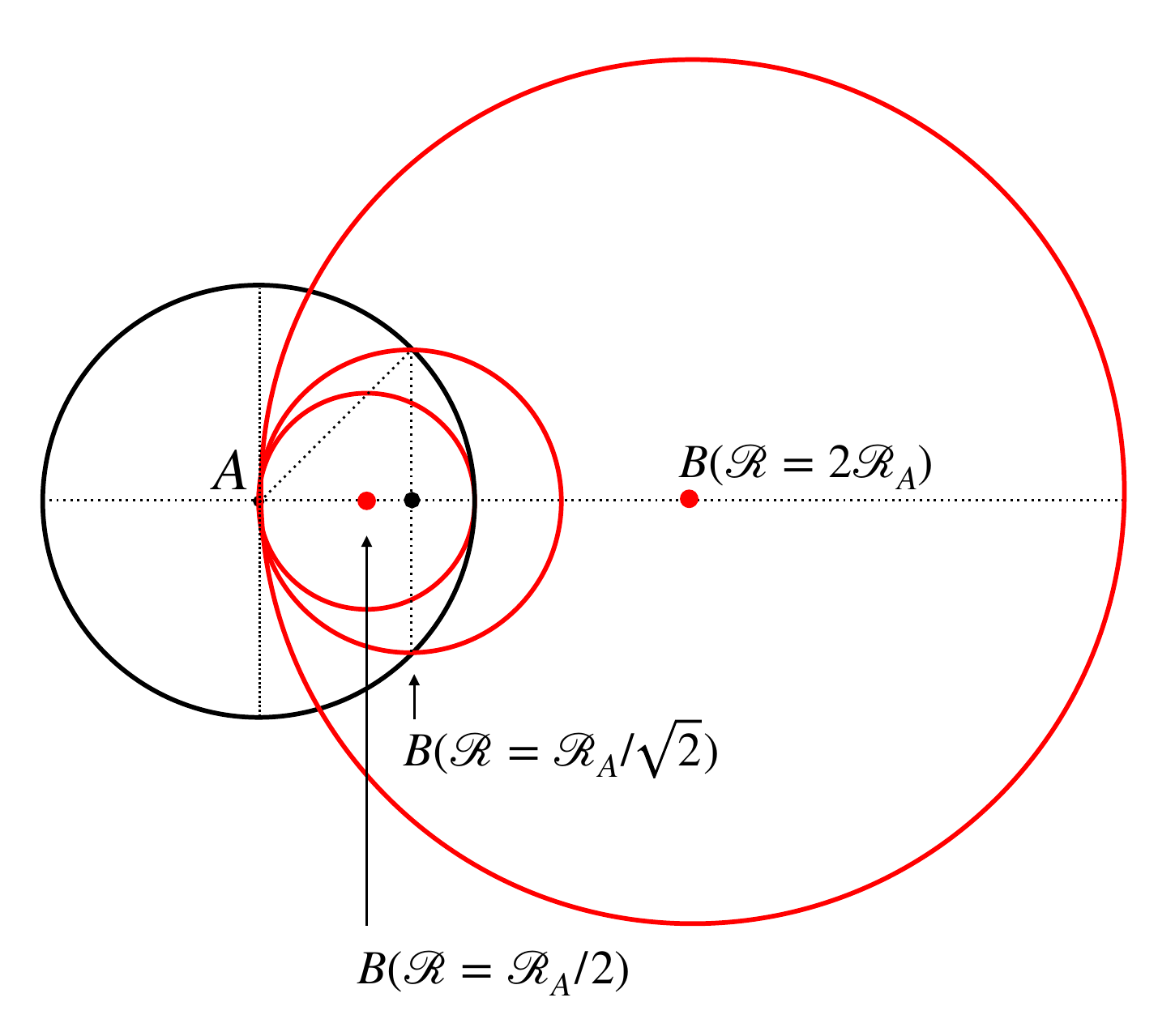}
\par\end{centering}
\protect\caption{Left panel shows a causal history, 
     with examples of horizons and footprints for  causal diamonds  that bound incoming null data at $A$ from  $B$ world lines along a particular direction.
 The right panel shows a 2D cross section of the same $A$ and $B$  spherical horizon footprints  at the end of inflation. 
 For the examples shown, ${\cal R}_B/{\cal R}_A= .5, 1/\sqrt{2}$ and $2$; a complete set would have $0.5< {\cal R}_B/{\cal R}_A< \infty$. 
Intersections of  $A$ and $B$ spherical footprints are circles  that represent  boundaries of coherent incoming null surfaces  that affect the CMB at the corresponding axial angle $\theta$, shown in Fig. (\ref{trig}).
 \label{ABcausal}}
\end{figure*}
\begin{figure*}[t]
\begin{centering}
\includegraphics[width=.45\linewidth]{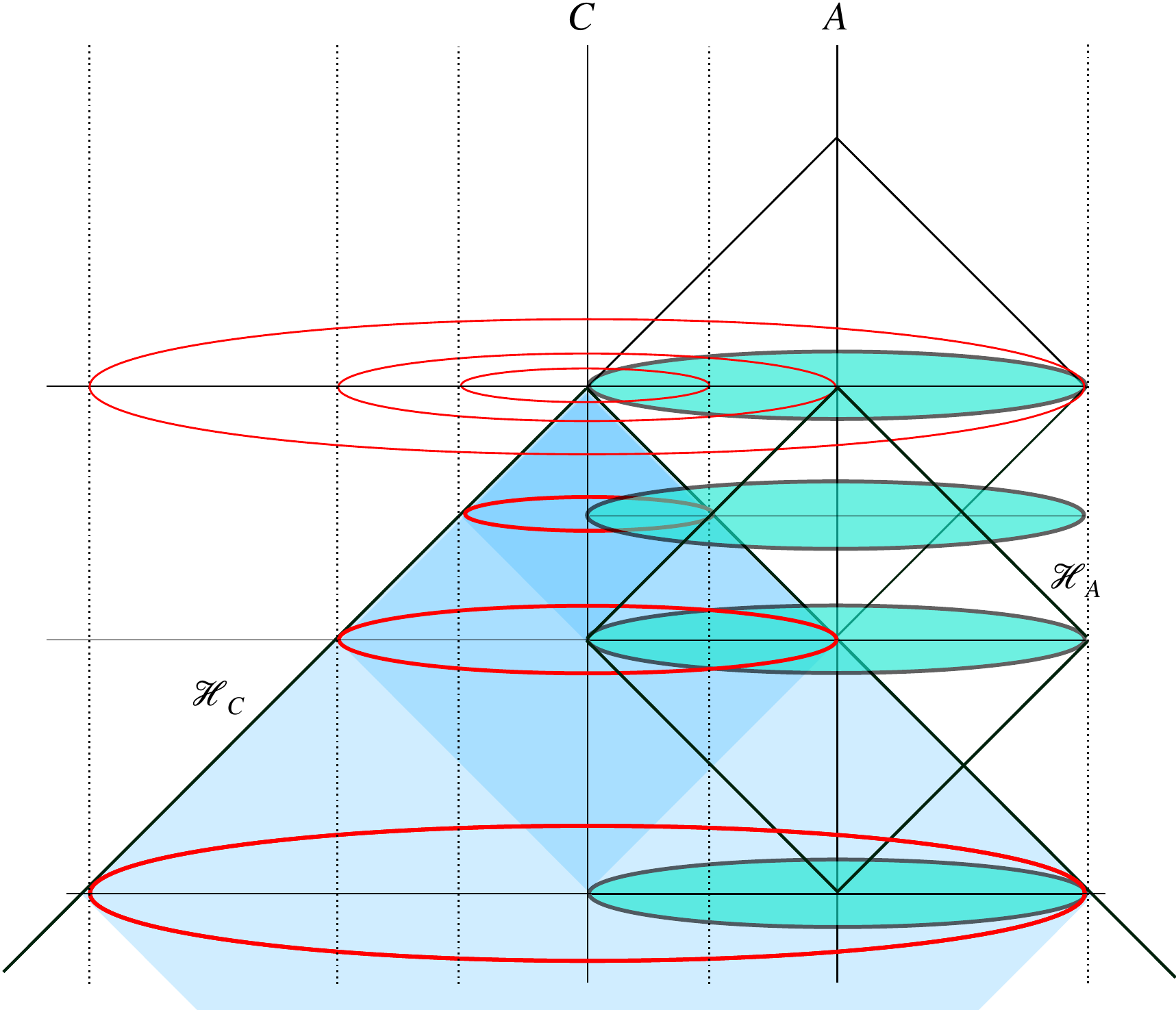}
\includegraphics[width=.49\linewidth]{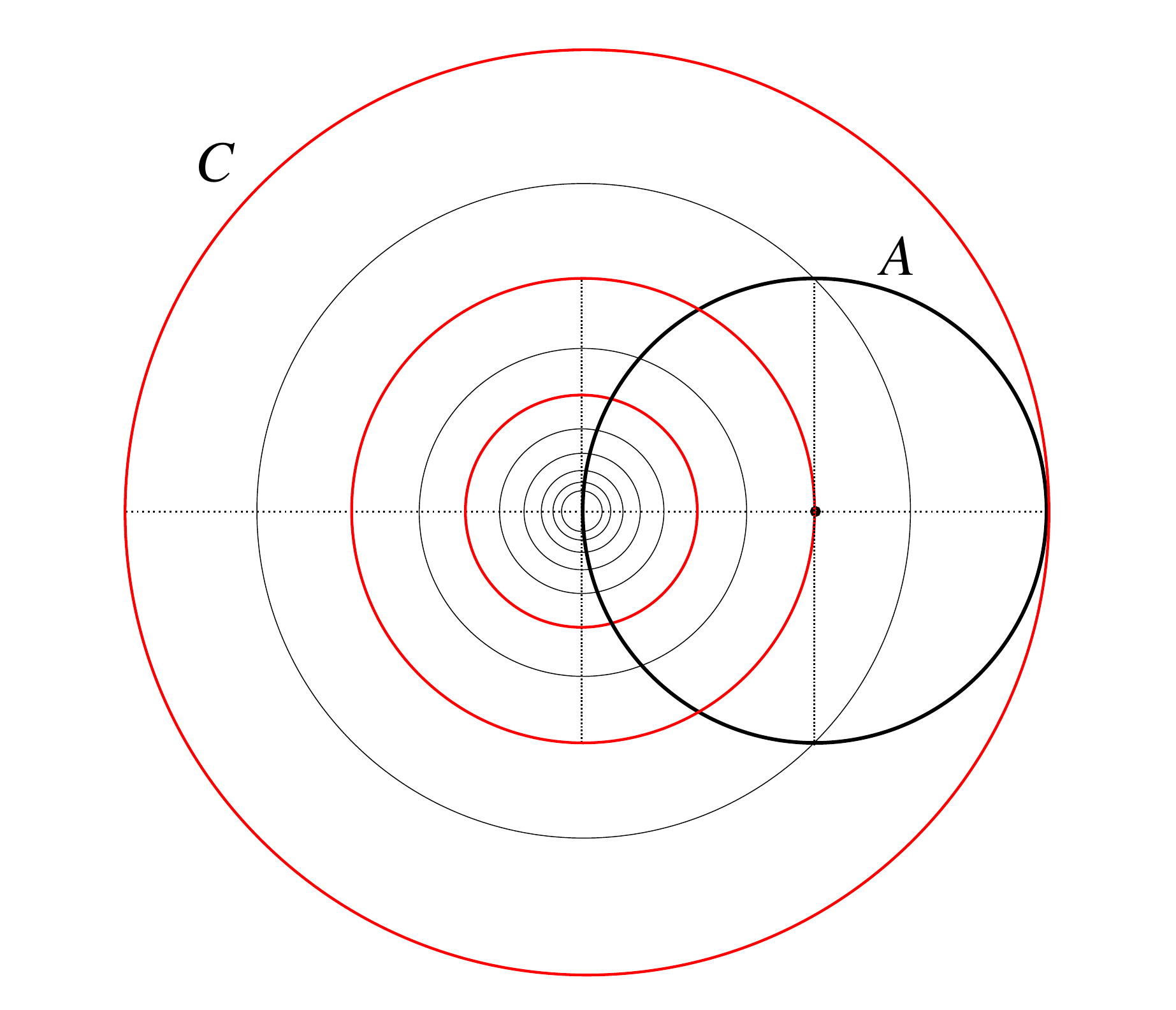}
\par\end{centering}
\protect\caption{Causal angular relationships  for causal diamonds  with outgoing null data from $A$ to a point $C$ on its  horizon. 
 At left,  the inflationary horizon of a world line $C$ that lies on $A$'s  horizon footprint, and footprints of some of $C$'s causal diamonds. 
 The right panel shows 2D cross sections of $A$ and $C$ horizon footprints  at the end of inflation. 
The series of $C$ footprints is shown on a linear scale, with constant logarithmic spacing in factors of $\sqrt{2}$. 
The largest shown here, with ${\cal R}_C= 2{\cal R}_A,$ is  the largest to causally entangle with angular correlations on the $A$ surface; the whole set has $0< {\cal R}_C/{\cal R}_A<2$. The sequence continues to smaller scales over the last $\sim 60$ $e$-foldings of inflation. 
 \label{ACcausal}}
\end{figure*}

In our model, the relic distortions at the end of inflation depend on intersections of causal diamonds and horizon footprints. 
 Figs. \ref{ABcausal} and \ref{ACcausal} show two classes of world lines 
 $B$ and $C$ whose inflationary horizons have particular causal relationships with an observer $A$: world line $A$ lies on the horizon footprints of $B$ world lines, and $C$ world lines lie on the horizon footprint of $A$.
Observable radial displacement is shaped by outgoing shocks from $A$ on $C$-centered causal diamonds, and incoming shocks to $A$ on $B$-centered causal diamonds.
 
\begin{figure}
\begin{centering}
\includegraphics[width=\linewidth]{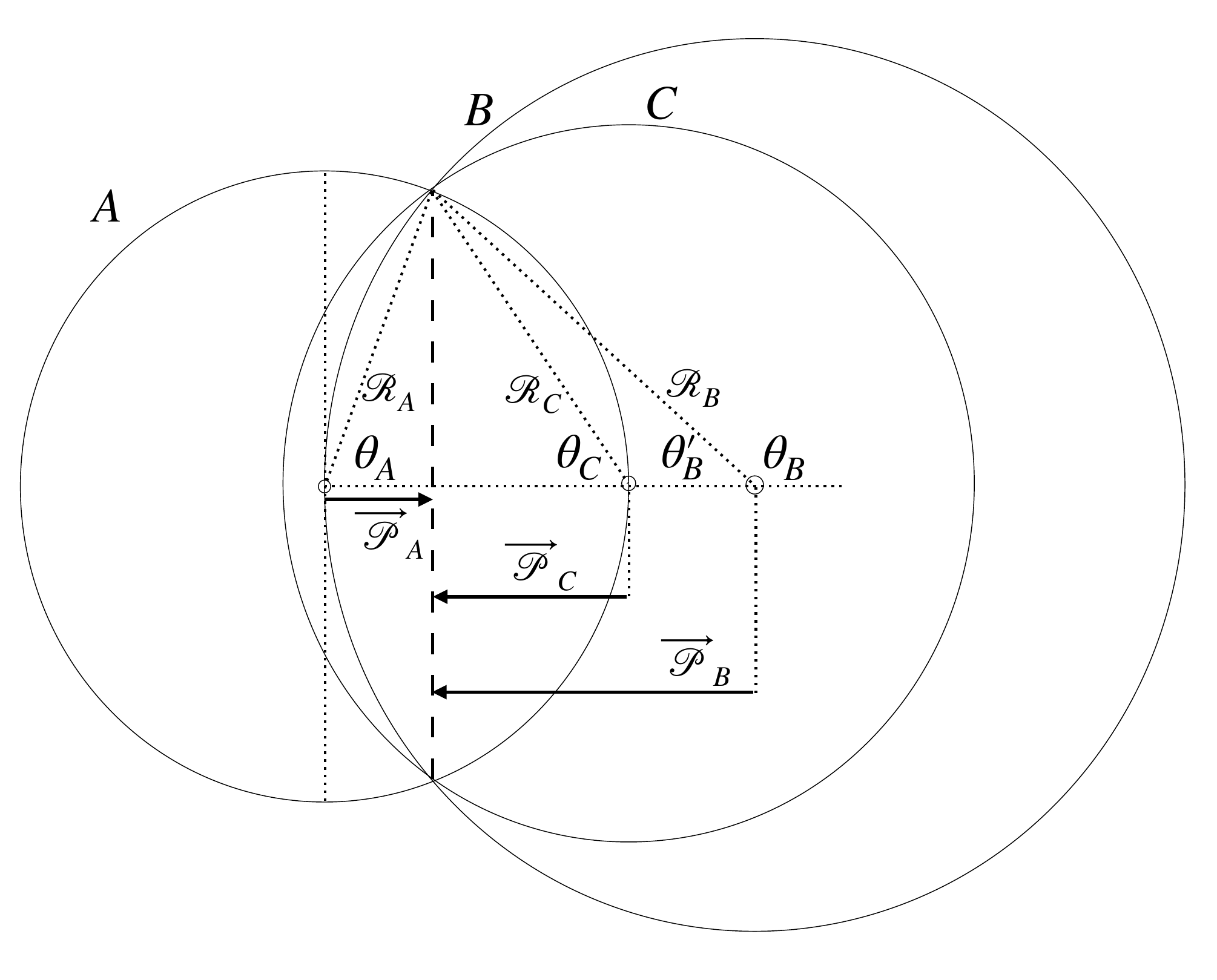}
\par\end{centering}
\protect\caption{Geometrical relationships between world lines and intersections of their spherical horizon footprints, shown in relation to their separation axis. Circles represent sections of three comoving spheres centered on three colinear points $A,B,$ and $C$. Angles $\theta_{A,B,C}$ (Eq. \ref{anglesABC}) and comoving radii ${\cal R}_{A,B,C}$ are shown for their common circular intersection, shown as a solid vertical line. An axially symmetric coherent directional displacement $d_A(\theta)$ on the $A$ horizon footprint is constant on the intersection circle. These angles with the common polar $ABC$ axis map onto boundaries of incoming and outgoing null cones shared by $A$ and its horizon footprint.
 \label{trig}}
\end{figure}

\begin{figure}
\begin{centering}
\includegraphics[width=\linewidth]{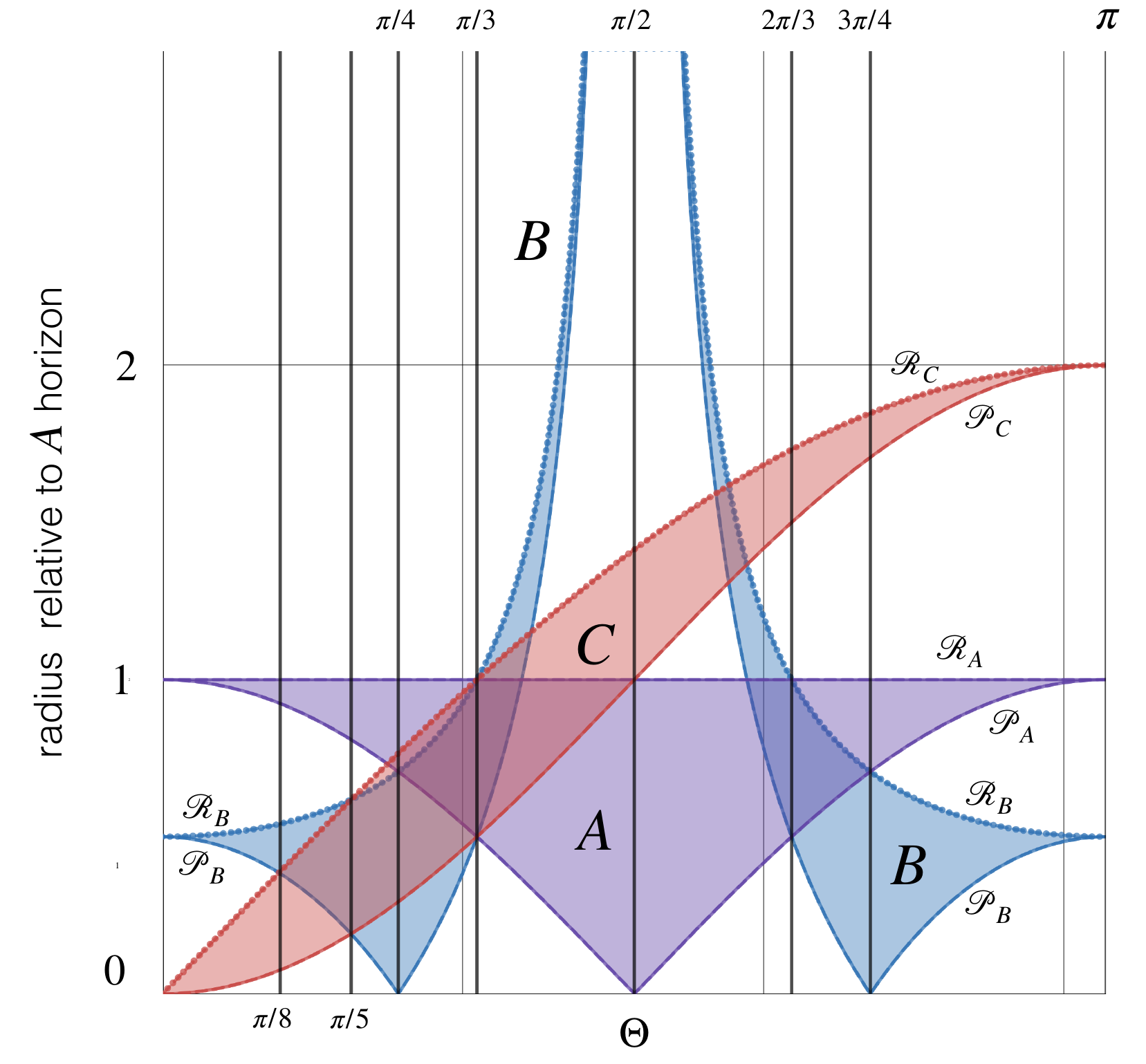}
\par\end{centering}
\protect\caption{ Comoving sizes of horizon radii ${\cal R}_{A,B,C}$, and polar-axis projections ${\cal P}_{A,B,C}$ to the intersection circle as a function of $\theta$, the $A$ polar angle. Length units are linear in units of ${\cal R}_A$. The time history of inflation runs from top to bottom. The ${\cal R}_{A,B,C}$ curves represent radial distances to the circles at polar angles where horizons intersect, and ${\cal P}_{A,B,C}$ curves represent distances along the polar axis to the intersection plane. 
In-common displacement of $A$ and $B$ causal diamonds leads to vanishing observable distortions on $A$'s horizon footprint between $\theta=\pi/4$ and
$\theta=3\pi/4$ from the axis, the boundaries of polar causal caps, where ${\cal P}_{B}$ vanishes. The two branches of $B$ spheres in opposite hemispheres have opposite parity.
\label{ABCradii}}
\end{figure}

The geometrical relationships are shown in cross-section in Fig. \ref{trig}. The polar angles of the horizon intersection from each center are:
\begin{equation}\label{anglesABC}
\theta_A=\theta,\ \theta_B=2\theta,\ \theta_C=(\pi-\theta)/2. 
\end{equation}
The time evolution of the comoving horizon radii, and separations from the intersection plane, are shown in Fig. \ref{ABCradii}.

\subsection{Model of measured displacement}

\subsubsection{Model of coherent axially symmetric displacements}

We now construct a causally coherent model of observable distortion. 
The aim is make a noise model whose elements are displacements on causal diamonds, constrained by  classical causal geometrical relationships  determined by the inflationary background.  The model is not a quantum theory; the goal  is to model  correlations expected for inflationary noise with causal coherence.

As in holographic models of black holes \cite{Hooft:2016cpw,Hooft2018,Hooft:2018syc}, and in the EPR example (Fig. \ref{clocks}), a causal diamond is assumed to be a coherent quantum system.
Noise from  null virtual shocks creates  coherent large-angle distortions of causal diamonds. Distortions of causal diamonds  relative to their centers are observable as anisotropy.

In the model,
vacuum noise is generated at a constant physical rate throughout inflation given roughly by Eq. (\ref{potentialvariance}). It creates perturbations of order $\Delta^2\approx 
d^2 H^2/c^2\approx Ht_P$ every Hubble time, where $H$ is the expansion rate and $t_P$ the Planck time. Distortions  are introduced as null virtual shocks on horizons throughout inflation, which represent  vacuum quantum fluctuations in position
 with zero mean and Planck variance.
Each shock leaves a memory,  a coherent relational displacement.

With this setup,
coherent displacements of world lines and their causal diamonds lead to correlated axially symmetric displacements on spacelike circles where spherical horizon footprints intersect. 
Every circle inherits anisotropic displacements by  quantum ``collapse'' of coherent causal diamond states. 
A virtual shock associated with a direction $\theta=0$ creates a correlated displacement in a circular ring described by a kernel
 $d(\theta)$, shaped by  radial projection of the coherent relative directional displacements of intersecting horizons $A$, $B$ and $C$ in the direction $\theta=0$.

The small-scale corrugation of $A$'s horizon footprint, the CMB  surface, continues  up to the end of inflation.
 Fluctuations between horizons in every direction continue to add successively smaller coherent perturbation patches as inflation proceeds, with quantum coherence over the inflationary horizon of each world line. 
Angular coherence is imposed both by displacements of the $A$ center, described as displacements on $B$ horizon footprints, and displacements of the $C$ center, described as displacement of $C$ horizon footprints. 

\subsubsection{Coherent displacement and causal shadow}
As shown in the classical EPR system, the physical effect of shock memory is a relational coherent displacement that refers to the difference of the center from the boundary of a causal diamond.
Coherent 
remembered displacements of whole causal diamonds do not produce observable anisotropy as viewed from the inside. 
For a classical homogeneous planar shock, a whole causal diamond is coherently displaced, meaning that the center and the boundary have the same displacement in any direction as viewed from very far away.

In the inflationary  system of overlapping shocks on horizons, in some directions,  displacement from shocks on other horizons is absorbed into the emergent position and velocity of the $A$ world line in relation to its horizon, and becomes unobservable. 
In  our system, the planar limit for $A$  corresponds to ${\cal R}_B\rightarrow \infty$, and an intersection at polar angle $\theta= \pi/2$. The planar limit for $B$ corresponds to $\theta= \pi/4$ and $\theta= 3 \pi/4$.

In our  model of coherent displacement, 
{\it observable displacement vanishes when the $B$ center is far enough away from the $A$ center that the $A$ center lies in the same hemisphere of $B$ as the $AB$ intersection circle.}
There is  a complete blackout of observable displacement, a ``causal shadow'', over a wide range of polar angles:
\begin{equation}\label{around90}
 d(\pi/4<\theta<3\pi/4) = 0.
\end{equation}
On the boundary of  this range, the intersection circle is a great circle on the $B$ horizon.
 Within   $\pi/4<\theta<3\pi/4$, the $B$ diamond boundary (which includes worldline $A$) shares a common displacement with the $A$ horizon footprint, and
there is no measured displacement.

This symmetry  
follows from a deeper physical hypothesis about the effect of coherent virtual  fluctuations in a space-time that emerges  from coherent quantum states on horizons:
that {\it distortion from  virtual fluctuations associated with any direction only comes from relationships between world lines in the same hemisphere as their horizon intersection.}
{ Conversely, virtual physical displacement  parallel to any direction between  events  only becomes measurable when the events
lie in opposite hemispheres of the  null cone that entangles them both with that direction}. In the application to CMB anisotropy, the displacement refers to the projection along any axial direction from  world line $A$,  relative to  events  in a circle  around that axis on the CMB surface. Measurable displacements occur when $A$ and $B$ world lines lie on the same side of the horizon intersection plane;  the causal shadow, with no displacement between the $A$ center and the circle, happens when $A$ and $B$ world lines lie on opposite sides of the intersection plane.

Physically, this symmetry relies on  exact balance of mean displacement in any two opposite hemispheres on the surfaces of a causal diamond---   a  generalized and delocalized momentum conservation or translation invariance associated with components of displacements parallel to any specific direction.
In an emergent causally coherent model, all points inside a causal diamond share the same mean relationships 
 of the diamond with the outside, so
  fluctuations of a point relative to the diamond boundary in any direction  average to zero over the whole spherical surface. 
  
  
  
  Causal shadowing is    consistent with a relational holographic algorithm for the emergence of macroscopic systems. At a given time  (that is, on a spacelike surface of constant time defined by any horizon footprint), {\it the pattern on the horizon surface at angular separation greater than $\pi/4$ from any direction has  not yet been affected by incoming null information from that direction}. This causal decoupling is the reason for the shadow in correlation, and also implies that  an existing reduced relationship at one radius is consistent with subsequent new information from farther away; subsequent information creates different patterns on larger  $A$ horizon footprints that become visible later. Similarly, outgoing information from $A$ that has not yet reached  distant world lines will not contradict the patterns they already see. The shadow thus preserves the independence of orthogonal directions in the emergent classical system. Relationships with  world lines separated by more than twice the comoving radius to a particular footprint (the conformal proper duration of an interval that determines its causal diamond) remain in an indeterminate, entangled state relative to that footprint.


At $\theta>3\pi/4$, for consistency the model requires  a ``print-through'' effect in one hemisphere, created  by   subtraction  of the coherent $B$ displacement pattern. 
 As in EPR,  this ``spooky'' effect is nonlocal but not  acausal, because its   state is prepared within  the overlapping $A$ and $B$ causal diamonds.  It is necessary for consistency: it represents the subtraction of the  displacement visible on $A$ that has been ``absorbed by'' coherent displacment of  $B$ horizons in one direction. 
 A nontrivial symmetry used in our model is that the removed pattern as observed internally on $B$ horizons should be the same as that on $A$ horizons, with the projective relationship $\theta_B= 2\theta_A$. 

In our model, the causal shadow  is the main  reason that causal coherence leads to relatively small correlations at large angles.

\subsubsection{Projection function $d_{AB}(\theta)$}

With the shadowing principle in mind, let us introduce a  projection function $d_{AB}(\theta)$  to
represent  the fraction of  displacement along the polar  $ABC$   shock axis 
that appears as the  radial displacement of a horizon footprint viewed from the $A$ center at an angle $\theta$ from the axis.
The radial displacement  is interpreted in  the emerged perturbed classical metric as a scalar perturbation of potential between $A$ and points on the spherical surface of a classical causal diamond, or as a distortion of causal structure. This part of relational projection does not depend on background curvature, only on the conformal causal structure.

We posit  the following simple function to describe  projection of axial displacement onto radial displacment:
\begin{equation}\label{simpleAB}
d_{AB}(\theta)= \cos (\theta)\cos(2\theta)
\left[\cos(\theta)-\frac{\sin(2\theta)}{{2}}\right].
\end{equation}


 In Eq. (\ref{simpleAB}),  the $\cos(\theta) $ and $ \cos(2\theta)$ factors  represent conventional geometrical projections of  axial  displacement onto radial displacements on the curved surfaces of $A$ and $B$ causal diamonds.
They can be interpreted as  the projections of axial shock displacements onto the radial measurement from $A$ (that is, the CMB horizon footprint), and onto the $B$ sphere surface that passes through $A$ and its footprint.
The onset of causal shadowing happens where the latter term vanishes, at $d_{AB}(\pi/4)=0$.

The expression in square brackets   is a smooth monotonic function added to mimic the  effect of null shocks on  causal diamonds with a relational causal shadow. It is approximately linear near its maxima and minima $\pm 1$ at the poles, $\theta=0$ and $\theta=\pi$. It has zero value and derivative at $\theta=\pi/2$, the  limit of  an infinitely distant $B$ sphere where a planar shock passes through $A$.

The overall product in Eq. (\ref{simpleAB}) approximates the response expected for causal shadowing: it is unity at the poles, and decreases to a much smaller value over the range of null causal symmetry (Eq.\ref{around90}).
At the precision of the current data, the final power spectrum and correlation function are not sensitive to the exact shape adopted for this model projection.

\subsubsection{Axial displacement $P_{ABC}(\theta)$ from scale-invariant inflationary  horizons}

The projection function (Eq. \ref{simpleAB}) is the part of the  model that describes the response of observed displacement to shocks from a particular direction. 
The other main element  is a  model of the  accumulated noise from null shocks in each direction over the specific history of slow-roll inflation.

In the inflationary context, the total displacement along the polar axis represents the cumulative effect of virtual Planck scale  displacements from  smaller and smaller angular scales up to the end of inflation. 
 The frequency of displacement noise is approximately constant in physical time, so the total displacement increases at small angles.
We  first describe a model that reproduces expected scaling of  holographic axial displacement in a scale-invariant cosmology, a de Sitter-like background with uniform spacetime curvature, 
and later add the effect of broken scale invariance or tilt as a correction.

The model invokes a  physical assumption  familiar from the phenomenon of freezing assumed in standard quantum inflation: after a horizon footprint becomes frozen into the expanding background, a virtual displacement of a definite physical length  stretches along with the classical comoving background.

Denote by $P_{ABC}(\theta)$ the full variation of displacement along the $ABC$ axis, before accounting for the effects of directional projection and cancellation described by $d_{AB}(\theta)$.
Our model for this is not based on a rigorously derived classical shock solution, but is an approximation constrained to match standard inflationary scaling of small-angle  perturbations  generated near the antipodes.

One contribution to distortion at small angles  arises from distortion of the boundary of the $A$ horizon footprint at  $C$ horizon intersections up to the end of inflation.
The comoving radius of the $C$ horizon at each angle is 
\begin{equation}
 {\cal R}_{C}=2{\cal R}_A{\sin(\theta_A/2)}. \end{equation}
(see Fig. \ref{trig}).
Since the inflationary horizons have a nearly fixed physical radius during slow roll (apart from a tilt factor added below),
 the redshift or ratio of scale factors corresponding to the ratio of the $A$ and $C$ horizon footprints is
\begin{equation}\label{CAredshift}
 a_C/a_A= {\cal R}_A/{\cal R}_C =\frac{1}{2\sin(\theta/2)}
\end{equation}
(see Fig. \ref{ABCradii}).
On the comoving  $A$ footprint outside ${\cal H}_A$--- that is, during inflation but outside the causally coherent horizon---  the $AC$ displacement in the axial direction stretches like other spacelike separations, so the same factor determines the $AC$ part of the total displacement: \begin{equation}
 P_{AC}= \frac{1}{2\sin(\theta/2)}. 
\end{equation}
 
A similar contribution to small angle perturbations comes from the intersection of $A$ and $B$ horizons close to both poles.
We adopt a similar expression that also gives the correct  inflationary scaling at small angles, but contributes symmetrically in both hemispheres:
\begin{equation}
 P_{AB}= \frac{1}{2\sin(\theta)}. 
\end{equation}
This displacement  arises from the two antihemispheric $AB$ differential relationships.

As in the projection function, our  model assumes that the total physical displacement in the emerged system is a difference between  these two effects.  This produces 
a total axial displacement $P_{ABC}$ that smoothly interpolates at all angles between  correctly-scaled logarithmic divergences near the antipodes:
\begin{equation}
P_{ABC}(\theta)\propto P_{AB} - \sin(\pi/4) P_{AC}.
\end{equation}
 A coefficient 
$\sin(\pi/4)=1/\sqrt{2}$ has been added so that the function vanishes at 
$\theta= \pi/2$.
This   symmetry is expected from the emergence of the local cosmic rest frame, since  total perturbations from opposite directions along any axis  produce a zero total displacement at the center by definition.
The shape of the function is not symmetric around $\theta=\pi/2$, due to the asymmetry in the way perturbations are generated in antipodal directions. 

Absorbing the common factor of 2 into an overall normalization then leads to 
 \begin{equation}\label{Pdiffer4}
 P_{ABC}(\theta)= \frac{1}{\sin(\theta)}-\ \frac{1}{\sqrt{2}\sin(\theta/2)}. 
\end{equation}
Formally, this function runs from $P(0)= -\infty$ to $P(\pi)=+\infty$. The logarithmic divergence is not physical because slow-roll inflation has a finite number of $e$-foldings. In slow-roll inflation, the divergence is controlled by a departure from scale invariance (the ``tilt''), as described below.

\subsubsection{Agreement with standard inflation on small scales}

Slow-roll inflation produces approximately equal contributions to variance in potential from each $e$-folding of inflation.
From Eq. \ref{potentialvariance}, the total variance is the logarithmic integral over the scale factor. For example, the $C$ contribution can be written in terms of ${\cal R}_C$ and hence $\theta$:
\begin{equation}
\frac{\langle\Delta^2\rangle_{>\theta} }
{Ht_P} = \int^{{\cal R}_C(\theta)} d\log( {\cal R}_C)= \frac{1}{\sin(\theta/2)}-1 . 
\end{equation}

Here, the constant of integration has been set so there is no contribution from $\theta>\pi$.
There are no pre-existing, longer wavelength modes, as there are in the standard theory: causal noise is associated with fluctuations within each causal diamond, whose comoving boundary represents the observed horizon footprint. 
The largest $C$ footprint involved in observable causal displacements corresponds to 
${\cal R}_C= 2 {\cal R}_A$.
Similar remarks apply for the $B$ horizon contributions; in that case, the smallest relevant $B$ footprints, which map onto $\theta=0$ and $\theta=\pi$, have
${\cal R}_B=  {\cal R}_A/2$.

The overall scaling in Eq. (\ref{Pdiffer4}) therefore matches  that of the standard spectrum of scale-invariant 3D perturbations in the small-angle planar limit,
at small angles 
($\ell>> 1$) where horizon curvature is not important.
With spectral tilt included as discussed in the following, the model agrees with the successful standard expectation
on small scales.

\subsubsection{Axial displacement with broken scale invariance (``tilt'')}
A correction for spectral tilt is needed to extrapolate to high $\ell$, or equivalently, to later stages of inflation. This factor accounts for the time evolution of curvature during slow roll inflation--- the fact that the physical radius of the inflationary horizon gradually increases.

One simple formulation of tilt as a function of angle is
\begin{equation}\label{tiltdefine}
 {\cal T}(\theta)=
 1-\frac{\epsilon}{2} \ln\left[P_{AC} \right]
 = 1+\frac{\epsilon}{2} \ln\left[2\sin(\theta/2) \right].
\end{equation}

 This factor accounts for the departure from scale invariance in slow roll inflation according to the scaling of horizon size: the physical size of the horizon $c/H$ increases slowly during inflation, and the curvature perturbation on horizons slowly decreases, by a few percent with each $e$-fold increase of the scale factor. If perturbations at each polar angle are dominated by horizons at scale factors proportional to 
 $P_{AC}(\theta)$, it leads to the form in Eq. (\ref{tiltdefine}). The tilt correction factor actually vanishes at a finite value of $\sin(\theta)>0$, so it  eliminates the formal logarithmic divergence of the scale-invariant theory at very small angles.

For a tilt that does not change with scale, the coefficient $\epsilon$ can be compared to a fit to perturbations at smaller angular scales measured by {\sl Planck}.
In the Planck data, the tilt for the 3D spectrum is $\epsilon_P= 0.035\pm .004$. The extra factor of two in the coefficient here accounts for the difference in tilt between power and amplitude.

A more physically accurate model (which in practice is almost indistinguishable from the one just described) includes separate tilt factors for the two terms $P_{AB}$ and $P_{AC}$, to account for different projections of inflationary history in antipodal directions. 
Define
\begin{equation}\label{tiltdefineAB}
 {\cal T}_{AB}(\theta)=
 1-\frac{\epsilon}{2} \ln\left[P_{AB} \right]
 = 1+\frac{\epsilon}{2} \ln\left[2\sin{\theta} \right]
\end{equation}
and
\begin{equation}\label{tiltdefineAC}
 {\cal T}_{AC}(\theta)=
 1-\frac{\epsilon}{2} \ln\left[P_{AC} \right]
 = 1+\frac{\epsilon}{2} \ln\left[2\sin(\theta/2) \right].
\end{equation}
Then, instead of the scale-invariant displacement factor $P_{ABC}(\theta)$ (Eq. \ref{Pdiffer4}), we define an axial displacement factor that includes tilt:
\begin{equation}\label{Ptilt}
 P_{ABC{\cal T}}(\theta)=
 P_{AB}(\theta){\cal T}_{AB}(\theta)-
 P_{AC}(\theta){\cal T}_{AC}(\theta){\cal T}_0/\sqrt{2}
\end{equation}
where
\begin{equation}
{\cal T}_0= \frac{{\cal T}_{AB}(\pi/2)}{{\cal T}_{AC}(\pi/2)}\simeq 1.006
\end{equation}
The coefficient ${\cal T}_0/\sqrt{2}$ preserves the symmetry described above that connects with the emergence of a local cosmic rest frame: the total displacement vanishes at $\theta=\pi/2$. Again, the shape of the function is not symmetric because the antipodal effects are different, but there is zero displacement in the normal plane.

\subsubsection{Total  displacement kernel $d_{\pm}(\theta)$}

We are now in a position to write down a  model for the total noise kernel $d(\theta)$, the  axially symmetric displacement on the horizon footprint of $A$ as a function of polar angle $\theta$ from a single shock axis.

Physically, the displacement kernel corresponds to the pattern of clock displacements everywhere on the $A$ horizon footprint produced by shocks originating from inflationary horizons along a direction $\theta=0$.
Although the choice of  direction on the axis is a matter of convention, the difference between them is physically significant, as discussed  below.

\begin{figure}
\begin{centering}
\includegraphics[width=\linewidth]{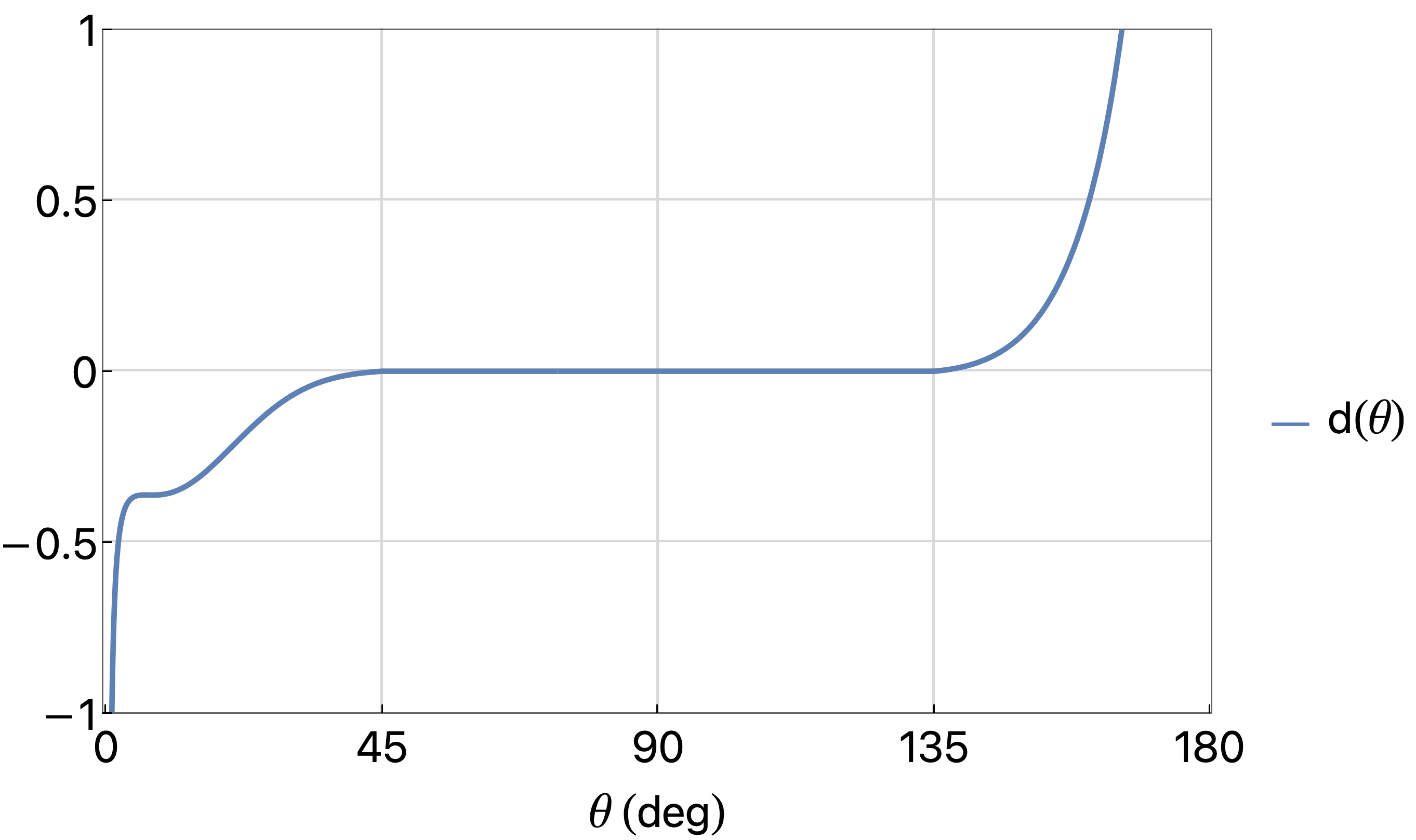}
\par\end{centering}
\protect\caption{Axially symmetric distortion kernel $d(\theta)$ (Eqs. \ref{polartilt},\ref{antitilt}) associated with a single polar direction.
The shape of the function has been derived from geometrical causal relationships but it has not been normalized to physical units.
The distortion identically vanishes for $45^\circ<\Theta<135^\circ$, the causal information shadow. The subtraction from in-common displacement of $A$ and $B$ causal diamonds appears at $\Theta< 45^\circ$ in this plot,
but the measured correlation is the same if the kernel is reflected around $90^\circ$.
 \label{mathdistortion}}
\end{figure}

The  model kernel is  constructed from a product of two factors  summarized above:  a projection factor
(Eq. \ref{simpleAB}) and an  axial displacement based on  inflationary background curvature 
(Eq. \ref{Ptilt}). 
The displacement $d(\theta)$ 
 takes a different form in the two polar caps $+$ and $-$ associated with the axis.
For the polar cap at $0<\theta<\pi/4$, 
\begin{equation}\label{polartilt}
 d_+(\theta)= P_{ABC{\cal T}}(\theta) \left[d_{AB}(\theta)- {\cal T}_{AB}(\theta)d_{AB}(2\theta)\right],
\end{equation}
and for the antipolar cap at
$3\pi/4<\theta<\pi$,
\begin{equation}\label{antitilt}
 d_-(\theta)= 
P_{ABC{\cal T}}(\theta) d_{AB}(\theta).
\end{equation}
The difference between the two hemispheres leads to parity violation in the resulting pattern (Fig. \ref{mathdistortion}). 

As discussed above, the subtracted $d_{AB}(2\theta)$ component in $d_+$  reflects  coherent displacement between the $A$ center and boundary  in one hemisphere.
This term represents an imprint of absorption in the $B$ pattern, the print-through of the coherent displacement of the $A$ center. 
This formulation is based on the symmetry that $A$ and $B$ horizons respond in the  same way to a displacement, as viewed from their respective centers. Thus, the function
 $d_{AB}$ is used with $\theta_B= 2\theta$ as an argument to represent the subtracted pattern on $A$ from internal $B$ horizon distortion.
 The observable distortion  in the two polar caps differs by
 the $B$ response, that is, $d_{AB}(2\theta)$, with a coefficient ${\cal T}_{AB}(\theta)$ to account for tilt. 

Since this model is  based on  geometrical relationships (including some approximate interpolations) rather than a complete physical theory, it would be reasonable to explore a wider range of options. In principle,  some  model elements  not  constrained by symmetries could be modified to yield better fits to  data. Given the limitations of current data, in this paper we chose instead to evaluate  consequences of this zero-parameter model and compare it with the standard model, in order to analyze the main new effects of causal coherence on observable anisotropy.

\section{Universal Angular Power Spectrum and  Correlation}

Up to a normalization, the kernel represented by Eqs. \ref{polartilt} and \ref{antitilt}
determines a unique angular power spectrum.
As a first approximation, it can be directly compared with CMB data in the Sachs-Wolfe regime, on the largest angular scales.
It has no parameters except for a normalization and tilt, which can both be estimated by standard fits to anisotropy on smaller angular scales. 

In our  model of perturbations, a realization of a sky pattern is a sum of $N>>1$ axially symmetric distortions from shocks along different axes uniformly distributed on the sphere, each with a  random amplitude. 
That is, the distortion of any horizon footprint from a uniform sphere in direction $\vec \Omega$ takes the form
\begin{equation}
    \delta \tau (\vec \Omega)= \sum_i^N \delta_i d(\theta= |\vec \Omega- \vec \Omega_i|) /c
\end{equation}
where $ \delta_i$ represents a random variable with zero mean for a shock associated with a direction $\vec \Omega_i$.  

In a holographic theory of gravity, the  variance of the displacement noise depends on the physical size of inflationary horizons, as in Eqs. (\ref{displacementvariance}, \ref{potentialvariance}). 
  For the purposes of this paper,  which addresses just the nature of the pattern at large angles, the normalization is arbitrary. 
  

Not all skies are the same, but in the $N\rightarrow \infty$ limit they all have the same angular power spectrum and  correlation function.
Up to a normalization,   we can compute the angular power spectrum and correlation function   of any realization just from the displacement kernel $d(\theta)$.
In standard notation (e.g. \cite{Hogan_2022}),
the anisotropic pattern on a sphere is decomposed into spherical harmonics
$Y_{\ell m}(\theta,\phi)$:
\begin{equation}\label{decompose}
\Delta(\theta,\phi)=
\sum_\ell \sum_m Y_{\ell m}(\theta,\phi) a_{\ell m}.
\end{equation}
The harmonic coefficients $a_{\ell m}$ then determine
the angular power spectrum:
\begin{equation}\label{powerpiece}
C_\ell= \frac{1}{2\ell+1}
\sum_{m= -\ell}^{m=+\ell} | a_{\ell m}|^2.
\end{equation} 
The angular correlation function is given by its Legendre transform,
\begin{equation}\label{harmonicsum}
 C(\Theta) = \frac{1}{4\pi}\sum_\ell (2\ell +1) { C}_\ell P_\ell (\cos \Theta).
\end{equation}

As in the EPR example, a single axially symmetric distortion reduces to one dimension, with $m=0$ in all the harmonics. 
The harmonic coefficients are then given by
\begin{equation}
 a_{\ell 0} = 2\pi \int_0^{\pi} d\theta \,\sin (\theta) Y_{\ell 0}^* (\theta)d(\theta).
\end{equation}

An impulse that is not aligned with the coordinate axis will produce different coefficients. 
However, the variance of the sum of  $a_{\ell m}$'s from many impulses is the sum of their variances;  a sum of  random impulses from $N>>1$ directions produces a  power spectrum that approaches the summed variance of power from each impulse,
 so any realization approaches  the same $C_\ell$ and $C(\Theta)$.
The model thus yields a definite angular power spectrum and correlation function  with no parameters except the    normalization. 
The difference between different realizations of sky patterns lies just in the phase factors of the $a_{\ell m}$'s.

The holographic model is  intrinsically more predictive than  the standard inflation quantum model, which produces many different realizations that have different power spectra on large angular scales. 
In the holographic view, almost all of the standard realizations are unphysical, because they cannot be generated by causally coherent displacements.
A holographic picture has fewer degrees of freedom, and the correlations of the angular pattern can preserve universal causal symmetries.

\begin{figure*}
\begin{centering}
=\includegraphics[width=\linewidth]{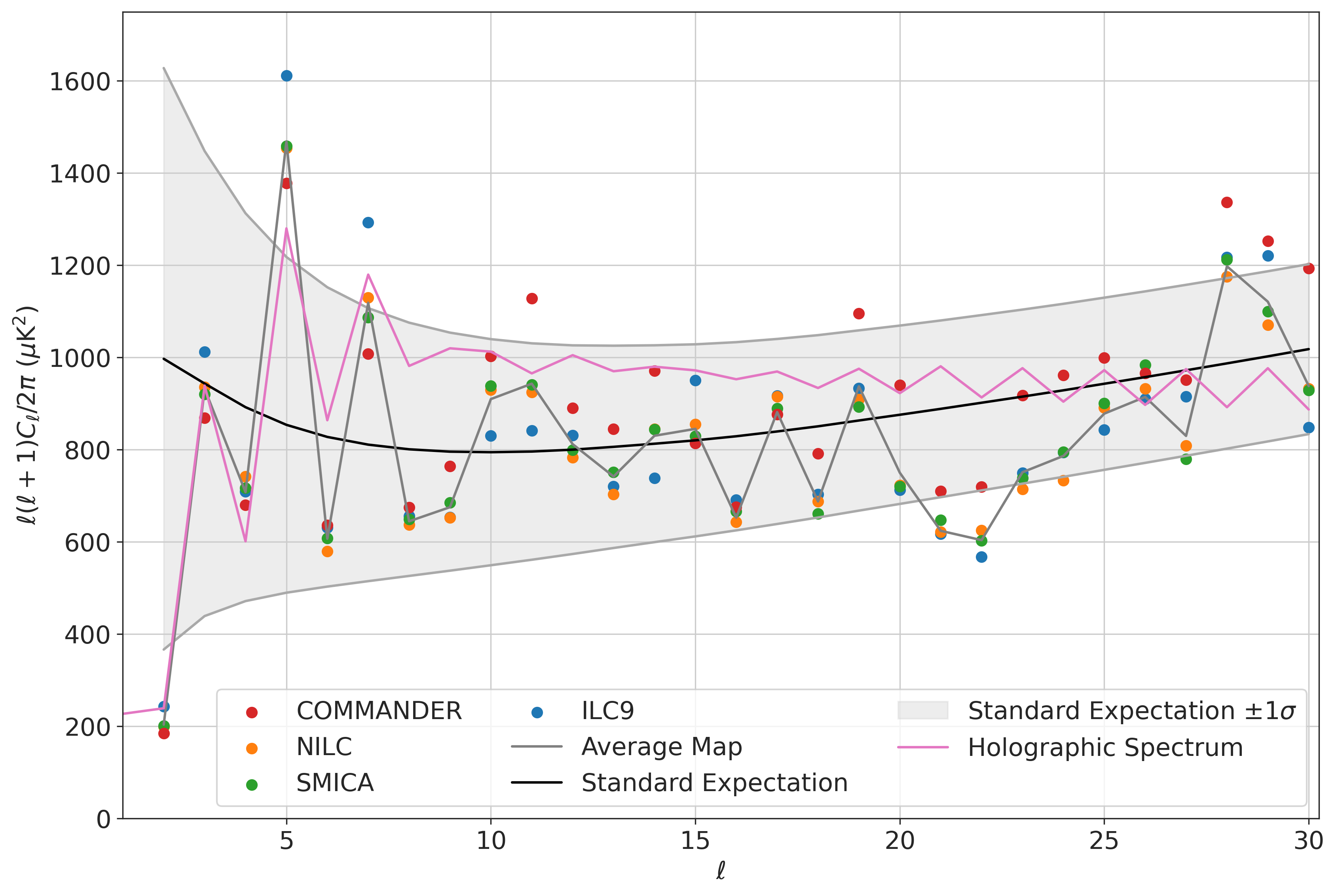}
\par\end{centering}
\protect\caption{Power spectra of models and Galaxy-subtracted maps for $\ell \leq 30$. The standard cosmological  expectation and cosmic variance band are shown for  {\sl Planck} fits to  cosmological parameters\cite{planckparams2018}. The holographic spectrum is the computed model Sachs-Wolfe anisotropy on a thin sphere,  fit to  the data with one parameter, a normalization,  by minimizing the quantity $\sum_{\ell = 2}^7 (C_{\ell, H} - C_{\ell, A})^2$. The holographic model has no cosmic variance, and departs significantly from the standard expectation at  $\ell\lesssim 7$, with a significant bias towards odd harmonics.
At $8 \lesssim \ell < 30$,  the model approximates the standard expectation for pure SW anisotropy, modified by a sawtooth pattern with a small nonstandard preference for odd harmonics.  The broad-band average in this range of angles differs  from the standard  expectation shown due to known  physical effects  not included in the SW approximation, such as scattering from reionization at $z\lesssim 20$, Doppler boosts, and ISW.
Residual error in maps due to models of Galactic emission, which dominates the scatter of points at each $\ell$, is a significant source of uncertainty that appears to be correlated at  different $\ell$.
 \label{holopowerspectrum}}
\end{figure*}

\section{Comparisons of models and data}

\subsection{Displacement and temperature distortions}

We have constructed a model for displacements $\delta\tau(\theta, \phi)$ of a causal diamond surface. In  the Sachs-Wolfe (SW) approximation\cite{1967ApJ...147...73S}, which is approximately valid on the largest angular scales, the same pattern appears in the temperature distortions of the CMB:
\begin{equation}\label{SW}
    \delta T_{SW}(\vec\Omega)\propto \delta\tau(\vec\Omega).
\end{equation}
We will assume this correspondence in the following comparisons. 
As discussed below, actual CMB temperature anisotropy depends on several known effects not included in the SW  approximation, but  included in the standard  model calculations.   We will compare models and data mainly on the largest angular scales,  $\ell\lesssim 7$, where
the  SW approximation works best.

The true  dipole distortion of our horizon ($\ell=1$) is unobservable, because of the degeneracy with the much larger dipole induced by our local  peculiar motion relative to the cosmic frame. 
Since the actual sky data has all of the the $a_{1m}$ harmonic components removed as part of the standard dipole subtraction,  comparisons of  $\ell=1$ components are not meaningful.
To compare real data with models of $C(\Theta)$, we  omit  the  predicted model dipole contribution.

In our interpretation, this dipole subtraction is the main reason the exact causal symmetry of true primordial angular correlation, which  vanishes over a wide range of angles, has remained hidden: the dipole needs to be included to obtain the true primordial correlation. 
 As shown below, if the predicted dipole is included, the model  displays the hidden causal symmetry.

\subsection{Datasets}

As in ref. \cite{Hogan_2022}, we use all-sky, foreground-subtracted maps prepared by 
 the {\sl Planck} team (Commander, SMICA and NILC) \cite{planckforegroundsub2018}, and the the {\sl WMAP} team, ILC9
\cite{2013ApJS..208...20B}.
For some tests, we also use a map generated by taking an average of these maps.

On the largest angular scales, the dominant uncertainty in tests of the holographic model arises from  Galaxy models, whose errors, unlike those of standard cosmic variance,   have significant and unknown angular correlations among different wavenumbers.  For some purposes it is possible to 
use the variation among the maps as a coarse measure of the uncertainty. We also note that there are significant differences in Commander between PR2 and PR3 Planck data releases on large angular scales, using the same Galaxy model, but updated datasets.
To test some aspects of our model, it may be possible to obtain more precise results using masks, but as shown previously,  care must be taken to avoid biased estimates\cite{Hagimoto_2020};  we defer this approach to future work.

Unlike most studies of the CMB power spectrum, we focus here on  $\ell\le 30$.
Variation among the datasets, dominated by different models of the Galaxy,  can be seen  in Fig. (\ref{holopowerspectrum}).
Some shared systematic variations in the spectrum, possibly  correlated at very different $\ell$,  could be an artifact generated by in-common  features of different  Galaxy models. 

\subsection{Angular power spectrum}

\subsubsection{General comparison with the standard model}

As noted above, a direct comparison of the model with CMB anisotropy assumes  the Sachs-Wolfe (SW) approximation, where the temperature of the CMB is  proportional to the gravitational potential, so the primordial pattern of horizon displacements appears  directly on the sky. 
The approximation works best on the largest angular scales, which also preserve the clearest signatures of causal coherence on curved horizons, and therefore the greatest contrast with the standard  model.  Thus, in Fig. (\ref{holopowerspectrum}) the model is shown normalized to fit the angular power spectrum of the average CMB map at $\ell\le 7$.
In spite of  measurement uncertainties, generally speaking the holographic model  appears to  agree with  data better than the standard model  at the largest angles. 

The holographic  model is constructed to reproduce the standard inflationary 3D power spectrum of initial potential perturbations  on small scales, so
with the correct normalization it  should match the expected SW spectrum  of the standard cosmological model.
The angular power spectrum in Fig. (\ref{holopowerspectrum}) roughly confirms this: at $\ell\gtrsim 7$ the predicted holographic spectrum is approximately flat with a slight downward tilt, as expected for the SW anisotropy of the concordance slow-roll cosmology, with a normalization that   roughly agrees with the expected normalization of the standard model SW effect at  $ \ell\gtrsim 8$.

\subsubsection{Parity violation}
An important systematic difference from the standard model  is that the holographic model 
violates parity.
On a sphere, parity violation means perturbations in antipodal directions tend to have opposite signs. 
In the angular power spectrum, parity violation appears as a sawtooth pattern,
a  preference for odd  over even parity harmonics.

Fig. (\ref{holopowerspectrum}) shows large systematic parity violation: an   excess of odd over even  fluctuation noise power in the holographic model at the largest angular scales ($\ell\le 7$) that also appears in CMB maps.  The excess includes the famously low quadrupole, and continues with a large fractional contrast up to  $\ell \simeq 7$. The predicted fractional parity violation becomes relatively small at $\ell\ge 8$, but some systematic parity violation in the holographic model persists even up to $\ell\sim 30$. 

Significant spectral parity asymmetry has previously been detected in  CMB data 
up to $\ell\simeq 30$ \cite{WMAPanomalies,Ade:2015hxq,Akrami:2019bkn}, in excess of that expected from cosmic variance in the standard model. 
Thus,  anomalous parity violation on the sky is not confined only to the largest angular scales, but  includes high-frequency odd-parity fluctuation power. It is possible that  this effect is a  signature of a primordial universal holographic asymmetry.

It has previously been argued\cite{Hogan2019} that significant parity violation is a generic property of inflationary perturbations from causally coherent  quantum gravity, due to a universal asymmetry of displacements on incoming and outgoing surfaces of inflationary causal diamonds. It was argued there that the ultimate source is not  parity violation in field interactions, but  the  time asymmetry of the inflationary background.
The holographic model analyzed here provides a concrete example of this effect:   parity violation appears as  different coherent shock response kernels in opposite hemispheres, that originate as explained above from different frozen displacements of outgoing and incoming null surfaces.
In the model this asymmetry appears explicitly in  factors represented by differences, including the  print-through of the causal shadow.

It was also noted in ref. \cite{Hogan2019} that universal   primordial  parity violation can produce observable  signatures both in the CMB and in the large-scale 3D pattern of  linear density perturbations at late times. The latter effect is a relic of many independent horizons on smaller scales, all of which share the same universal parity-violating initial anisotropy.

 A universal primordial angular parity violation  as large as that in the universal holographic angular spectrum at $\ell\le 8$  could  in principle produce a significant relic parity violation in  3D  linear density perturbations on  sub-horizon scales\cite{Hogan2019}. We conjecture (but do not prove here)  that it may be large enough to account for the significant parity violation recently detected in 4-point correlations of the  galaxy distribution\cite{Cahn2021,Hou2022,Philcox2022}, without altering the directionally-averaged 3D power spectrum from the standard model.

\subsubsection{Corrections to the SW approximation}
After averaging over  the sawtooth pattern, at $7<\ell <30$ the holographic  model  still differs in detail from the standard expectation shown in Fig. (\ref{holopowerspectrum}). This difference is expected because several post-inflation  physical effects on temperature anisotropy, which are not included in the  holographic SW model, are not entirely negligible in this range of scales: for example,
 the integrated Sachs-Wolfe (ISW) effect from structure during the dark-energy-dominated era at $z\lesssim 1$, the Doppler effect of moving scattering  matter at $z\simeq 10^3$, and radiative transport, including scattering from reionized matter at $z<<10^3$.
 These effects also introduce a third dimension and thereby add a source of cosmic variance not present for thin spheres.
 
   More detailed modeling will be required for a consistent detailed match of  holographic initial conditions with classical post-inflation cosmology.
   As in the standard model, the Doppler contribution increases with  $\ell$, and increases steeply at $\ell\gtrsim 30$, where it likely overwhelms any holographic effect.
  The largest effect we have neglected in the range $7\lesssim \ell\lesssim 20$ is reionization: a scattering optical depth $\tau\simeq 0.1$ creates a ``fog'' that reduces $C_\ell$ by factor $\sim e^{-2\tau}\simeq .8$  on scales up to the horizon at reionization, and will thereby bring the relatively unblurred holographic spectrum at $\ell\lesssim 7$ into closer agreement with the standard  expectation at $7\lesssim \ell\lesssim 20$. Since the holographic model  implies different priors for low $\ell$  harmonics, it will  modify standard estimates of $\tau(z)$.

 \subsubsection{Comparison with other causally coherent systems}
 
We note that the low-$\ell$ angular spectrum of the causal distortion pattern differs significantly from noise spectra computed for models  of other causally-coherent gravitational systems, such as holographic  (or ``geontropic'') noise in interferometers (e.g., \cite{Li2022,McCuller2022}), or  a system of randomly oriented EPR-like shocks\cite{Mackewicz2022}. Part of the difference arises from the different space-times assumed for the causal structure of the background.  For example,    inflation creates  a nearly flat, slightly tilted spectrum at  high $\ell$, whereas holographic noise in flat space-time falls off as a power law at high $\ell$.  Differences between models  also arise at low $\ell$ from different
 sources of causally coherent noise; for example, the EPR source  has reflection symmetry, and produces purely even-parity modes. Different parity violation occurs from different symmetries in both the source and background.   However, all  of these cases exemplify similar distinctive nonstandard features of  causally coherent fluctuations:  the overall displacement variance is dominated by large angles and low-$\ell$ modes, with a large value given by a standard quantum limit for causally coherent noise (Eq. \ref{displacementvariance}).

\begin{figure*}
\begin{centering}
\includegraphics[width=\linewidth]{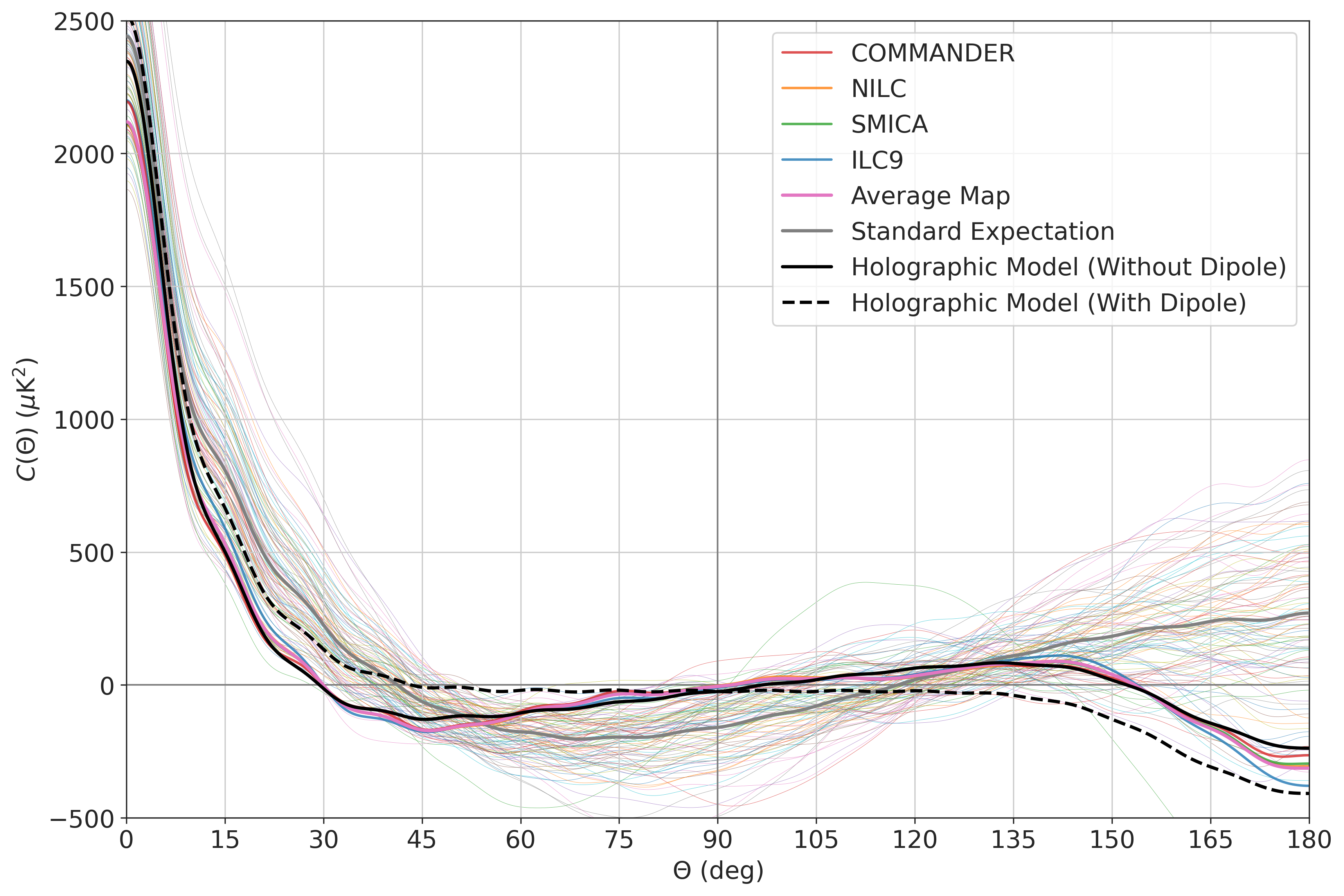}
\par\end{centering}
\protect\caption{  
Correlation functions $C(\Theta)$ predicted by models and measured in  maps, all shown with a low pass filter, $\ell\le 30$.  The standard expectation is shown for {\sl Planck} fitted parameters,  and  light colored lines show a corresponding random sample of 100  standard predicted realizations.
The holographic model normalization has  been set to minimize  $\sum_{\ell = 2}^7 (C_{\ell, H} - C_{\ell, A})^2$.
The model  shown with no dipole can be  compared directly with data on large angular scales.
The holographic model is also shown  with the predicted unobservable dipole included to display  hidden causal null symmetry (Eq. \ref{nullsymmetry}). 
 \label{correlationspider}}
\end{figure*}

\subsection{Angular correlation}

\subsubsection{General comparisons of models and maps}

Because correlations in the  angular domain  respect the sharply delineated  boundaries of primordial relationships among inflationary causal diamonds,  the angular correlation function $C(\Theta)$  preserves
 distinctive, direct signatures of causal coherence. 

In ref. \cite{Hogan_2022} we used CMB data to test conjectured null symmetries of  $C(\Theta)$. However, angular symmetries alone  are not sufficient to uniquely determine a power spectrum,  definite values of $C(\Theta)$ outside of a limited interval, or a definite amplitude for the intrinsic dipole.  
Here, with a definite  model that  yields a  full  function $C(\Theta)$ up to a normalization, we can do more powerful tests.

Figure (\ref{correlationspider}) shows the holographic   $C(\Theta)$ with and without the predicted dipole component, plotted using only harmonic components with $\ell\le 30$.
With the dipole omitted, the model $C(\Theta)$  can be compared directly on large angular scales with dipole-subtracted CMB maps,  and with realizations of the standard inflationary model. 
It is visually striking that the unique holographic model comes much closer to the data than any of the standard realizations.  Again, there are no
 adjustable shape parameters to achieve this agreement: the only parameter is the overall normalization, which is independently constrained to agree with the normalization of the standard expectation on small scales to be consistent with other cosmological tests.

The range of standard realizations illustrates typical cosmic variance. According to the holographic hypothesis, these realizations are unphysical:  the large-angle cosmic variance in the standard model results mainly from incorrectly calculated  long-wavelength fluctuations, whose causal coherence is not included in QFT\cite{HollandsWald2004, Stamp_2015}. 

\subsubsection{Model-independent  exact causal symmetries of $C(\Theta)$}

With the dipole included, an approximate primordial causal null symmetry is apparent in Fig. (\ref{correlationspider}):
\begin{equation}
    C(\pi/4\le \Theta\le 3\pi/4)\simeq 0.
\end{equation}
Although the causal distortion kernel of our model (Eqs. \ref{polartilt} and \ref{antitilt}) exactly vanishes in this range of angles by construction, the computed correlation does not exactly vanish at finite resolution.
In the limit of high resolution,  causal coherence   produces an exact  symmetry for the  angular correlation of  horizon distortion:
\begin{equation}\label{nullsymmetry}
   C(\pi/4\le \Theta\le 3\pi/4)_{\ell_{max}\rightarrow \infty}\rightarrow 0.
\end{equation}
In this limit, the dipole-subtracted SW anisotropy is then predicted to take an exact form
\begin{equation}\label{causalsymmetry}
   C(\pi/4\le \Theta\le 3\pi/4)=-c_1\cos(\Theta),
\end{equation}
with a coefficient $c_1>0$ that depends on the unmeasured  dipole. 

As discussed above, this  causal symmetry is independent of the details of the displacement model, as it originates directly from the boundaries of the causally connected causal diamonds--- the incoming and outgoing causal information that connects the observed horizon with our world line during inflation. 
The exact form (Eq. \ref{causalsymmetry}) predicted for dipole-subtracted primordial SW correlation   can in principle be used as the basis for direct,  general tests of primordial causal coherence.  
The  ``causal shadow'' conjectured here extends equally into both hemispheres,  over twice the range of angles $ \pi/2<\Theta< 3\pi/4$ that was previously conjectured and tested\cite{Hogan_2022}; this property simplifies model-independent tests, since even- and odd-parity symmetries apply independently, and the even component is independent of the unmeasured intrinsic dipole.  
With precise maps, these general properties can  be  adapted to model-independent tests of  causal coherence.
It was found in ref. \cite{Hogan_2022} that precise model-independent tests are currently limited by systematics of Galactic foreground models.


\begin{figure*}
\begin{centering}
\includegraphics[width=0.8\linewidth]{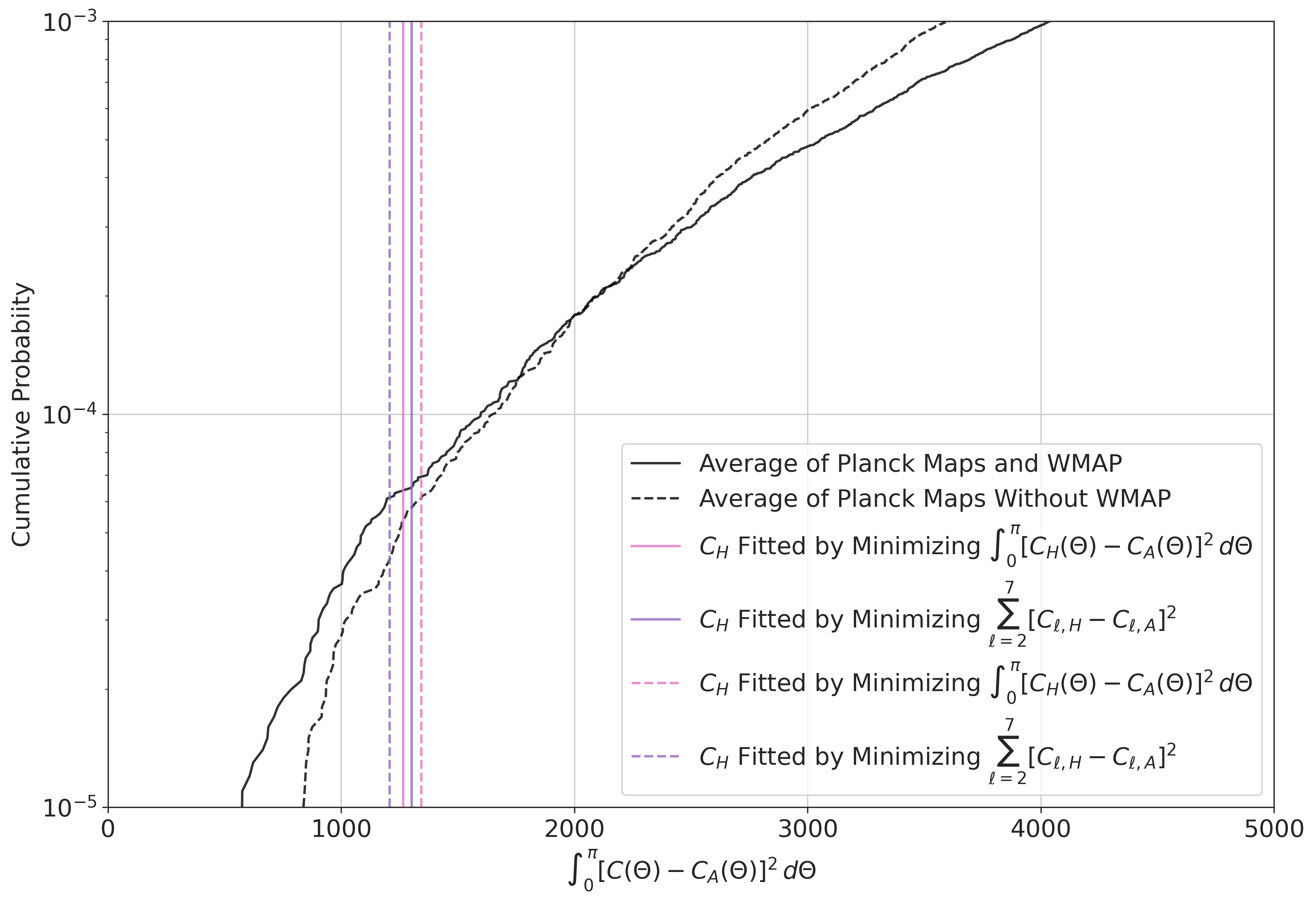}
\par\end{centering}
\protect\caption{Comparison of cumulative probabilities for the summed residual variance  $\delta C^2$, Eq. (\ref{residue}), from the measured  correlation function of average CMB maps, $C_A$.  The black solid and dashed curves show the fraction of standard-model realizations for which  $\delta C^2$  is less than the value on the horizontal axis.
By this ranking, fewer than 70 out of $ 10^{6}$ standard realizations come as close to the sky as the correlation function $C_H$ of the holographic model, whose residual $\delta C^2$ is shown by the vertical lines for several normalizations. This conclusion is not sensitive to model normalization or the choice of  map.
 \label{cumulative}}
\end{figure*}
\subsubsection{Comparison   of $C(\Theta)$  residuals for models and maps}

We now compare the relative likelihood of obtaining the measured sky according to two models, the  holographic model and the standard one.

The two models   have different priors for uncertainties on large angular scales. In the standard model, the dominant uncertainty
comes from cosmic variance, which introduces independent  random noise in realizations at each angular wavenumber $\ell$. In the holographic model, there is almost no cosmic variance on the largest angular scales: spectral powers at different $\ell$ are not independent, but have fixed ratios, and $C(\Theta)$ has a fixed shape.
In this case the dominant uncertainty, which is much smaller than standard cosmic variance on the largest scales,  comes from systematic errors in models of the Galaxy, which are also highly correlated among different $\ell$. The two models gradually converge at higher $\ell$, as  cosmic variance is added in the holographic model due to 3D effects such as scattering from a finite photosphere, and the  variance from the Doppler effect of radial motion.

In this context, we can still  perform a quantitative comparison of the relative likelihood of  obtaining the CMB maps in the two models,  using a method\cite{Hogan_2022} based on sums of residuals of $C(\Theta)$ to rank how well models approximate the data.

Let $C_M(\Theta)$ denote a model correlation function and let $C_D(\Theta)$ denote a data set correlation function. 
 Let $\{\Theta_j\}_{i = 1}^N$ be a uniformly spaced lattice of points in 
   the interval $[0, \pi]$.
 Then, define the residual
\begin{equation}\label{residue}
     \delta C^2 = \sum_{i = 1}^N [C_M(\Theta_i) - C_D(\Theta_i)]^2 \cdot \left(\frac{\pi }{N}\right),
\end{equation}
which is an approximation of the integral
\begin{equation}
    \delta C^2 = \int_0^\pi [C_M(\Theta) - C_D(\Theta)]^2 \, d\Theta.
\end{equation}
Smaller values of $\delta C^2$ indicate a better agreement. In practice, we found that $N = 2,500$ is a sufficiently high lattice resolution to approximate the integral among different data sets and models with negligible error.

We then evaluate the  $\delta C^2$  for many standard realizations of the standard model, such as those shown in Fig. (\ref{correlationspider}), as well as the holographic model. The comparisons is done for different maps to estimate the sensitivity to Galactic model uncertainties.

Fig. (\ref{cumulative})  shows 
 cumulative probability,  the fraction of standard realizations with $\delta C^2$ smaller than the value shown on the horizontal axis.
 For this comparison, all  maps and models are low-pass filtered at $\ell\le 7$.
 Two curves are shown: one for a comparison with a sky average of the three {\sl Planck} maps, and another with a sky average that also includes {\sl WMAP}.
The figure also shows $\delta C^2$ evaluated for the holographic model normalized in two ways to each of these maps, fitted to $C_\ell$  or $C(\Theta)$.
Note that the holographic model does not include  late-time scattering, a simplification that reduces $C(\Theta)$  at small  separation ($\Theta\lesssim 5^\circ$),  so this comparison is biased in favor of the standard  model.

In our simulated ensemble,  {\it fewer than 70 out of $10^6$ standard realizations fit the data as well as the holographic model.}
This comparative ranking  verifies the visual impression that the standard model almost never produces a correlation as close to the data as the holographic model.
We have found that this conclusion is not sensitive to the exact choice of normalization,  the choice of CMB map,
or the lower bound of angles used for the comparison.
If the holographic model is correct, the relative likelihood of the standard model should decrease further with less contaminated maps that more closely approximate the true primordial pattern.

\section{Conclusions}


We have presented a geometry-based  noise model  based on classical  projections of standard  causal relationships during slow-roll inflation, with displacements based on frozen-in geometrical memory on standard inflationary horizons. 
 The model avoids  well known  inconsistencies introduced by renormalized gravity in QFT  from entanglement with  acausally prepared long wavelength states.
 It describes  realizations of primordial  noise whose  causal coherence agrees with the causal structure  of  standard inflationary geometry.

Improving upon our earlier phenomenological model\cite{Hogan_2022}, the  model developed here 
 provides a quantitative (if still approximate) example of how geometrical causal   constraints and directional symmetries determine a correlation function $C(\Theta)$, including an intrinsic dipole amplitude.  On large scales it can be directly compared with dipole-subtracted data with no additional assumptions or parameters, apart from an overall normalization.
It has an exact  symmetry that can be traced to standard causal relationships of intersecting inflationary horizons: correlations of primordial  distortions on spherical surfaces at the end of inflation are predicted to vanish over a range of angular separation from $\pi/4$ to $3\pi/4$.

Within the estimated uncertainties, we find that the holographic model agrees in detail with the observed CMB correlation function on large angular scales. In particular it produces large-angle null symmetry and parity violation 
that seldom occur in the standard model.
The predictive power of the model contrasts with the standard model, where any particular outcome can in principle occur by chance, but very few realizations come as close as the holographic model does to the real CMB sky, even allowing for current uncertainties about  Galaxy models.
In our interpretation, the anomalously small correlation of the CMB at large angular separation signifies a fundamental symmetry of primordial perturbations.

Causal coherence can reconcile classical
 locality with  space-time relationships that  emerge from a delocalized quantum system.
``Spooky'' symmetries arise because a quantum system nonlocally entangles future and past events over whole causal diamonds. In a causally coherent cosmology, the quantum system is not reduced to a classical one until the end of inflation, even for long wavelength modes.
 The freezing or collapse of coherent causal-diamond quantum states  happens causally:
even now, as  we watch the sky,  the  quantum state of our causal diamond at the end of inflation is collapsing on its boundary to a definite, post-inflation classical causal relationship with our world-line, which causally entangles the observed values of temperature everywhere on the sky. Beyond that causal diamond surface, the exact causal structure of the ``background'' is still an indeterminate quantum system. Such indeterminacy follows directly from the causal coherence imposed to solve the IR inconsistency of quantum field theory.

A consistent holographic cosmology thus implies  much more than just a  correction to the standard angular spectrum on large angular scales at the end of inflation:  causally-coherent quantum gravity and emergent locality shape geometrical fluctuations throughout inflation, and  lead to radical
revision of much standard lore about quantum cosmological initial conditions and primordial symmetries. 
For example, in the absence of a definite infinite background on the largest scales,  even the standard classical decomposition of linear perturbations into scalar, vector and tensor components is not well defined; even in the post-inflation classical regime,  the  perturbative 
interpretation of a  measured geometrical relationship  depends on the time and location of the measurement. 
At the same time,  perturbations  on scales much smaller than the horizon  have the same broad-band, direction-averaged 3D power spectrum 
as the  standard model of slow-roll inflation, so the  successful match to small scale anisotropy, and comparisons with late time cosmic structure, are not substantially changed from the  standard model.

A notable departure from the standard picture on all scales is significant parity violation,   manifested in the CMB as antipodal anticorrelation $C(\Theta\rightarrow\pi)<0$ in the  correlation function, and a  bias to odd harmonics in the angular power spectrum.  
In a holographic model, parity violation in linear perturbations is predicted to be significant, universal and ubiquitous, and should be measurable today as significant parity violation in linear density perturbations  on sub-horizon scales\cite{Hogan2019}.  In principle, it is  large enough to account for  recent measurements of parity-asymmetric 4-point correlations in the large-scale linear galaxy distribution\cite{Hou2022,Philcox2022}.  

The  geometrical model advanced here is intended as an approximation that illustrates generic effects of causal coherence.
  All of the relationships used in the model are causal, but not all of the detailed choices made in its construction  have a rigorous justification, and some of them are clearly first approximations.
  Even so, because of the generic causal constraints used to build the model, we conjecture that its  correlations  share new and distinctive properties with  holographic perturbations that would actually be produced by emergent quantum gravity.
  General effects of causal coherence not present in the standard picture include universality and symmetry of $C(\Theta)$ at large angles exemplified by
   causal shadowing,  an axially symmetric directional displacement kernel associated with each direction, and a realization of universal parity violation.
The existence of a  geometrical causal model that accounts for  large-angle CMB correlations  is an indication that spacetime emerges from a causally and directionally coherent  quantum system.


 Even with a  universal primordial angular spectrum of causal distortion on horizons,    more precise holographic predictions for the CMB power spectrum, especially  at $\ell\gtrsim 7$ and/or including polarization,  will require modeling of  late-time physical effects beyond the SW approximation.
 Causal coherence should lead to new  predictions for higher order angular correlations on large angular scales, and may offer new ways to  account for  other apparently significant anomalies from standard predictions\cite{2017MNRAS.472.2410A,2015MNRAS.449.3458C,Selub_2022,Perivolaropoulos2021} not studied here. 
 
 Similar large-angle correlations  should  affect measurements of  any  coherent pattern of quantum geometrical distortions introduced on macroscopic causal boundaries.  The  approach adopted here, based on coherent spherical virtual null shock displacements, can be adapted to model and search for the effects of causally-coherent  noise  in  laboratory systems, including   interferometer experiments under development.
An important theoretical property of both our phenomenological cosmological holographic model and the EPR system is that coherent displacements are not scalars, but are directional and relational, in the sense that they refer to displacements on the boundaries of causal diamonds relative to their centers.
Directional coherence affects how  temporal statistics of holographic noise in an 
interferometer depend on the spatial arrangement of its arms\cite{Kwon2022}.

 The prediction of a falsifiable exact fundamental angular symmetry for large-angle primordial structure adds motivation to the quest for  precise maps of the true, uncontaminated primordial pattern on the largest angular scales. A satellite telescope\cite{Kogut_2011} with broad frequency coverage and angular resolution of a few degrees would help development of  accurate models for the Galaxy and hence  sharper probes of  primordial relic symmetries. 
 
\ 

\  

\begin{acknowledgments}
Useful comments and discussions were contributed by O. Kwon.
We are grateful for support from an anonymous donor.

\end{acknowledgments}

\bibliography{universal}

\end{document}